\documentclass[amsmath, amssymb, aps, eqsecnum, longbibliography,
  natbib, prd, reprint, superscriptaddress, twocolumn,
  floatfix]{revtex4-1}
\usepackage[dvipdfmx]{graphicx}
\usepackage[dvipdfmx]{hyperref}
\def\aap{Astron.\ Astrophys.\ }
\def\apjl{Astrophys.\ J.\ }

\def\mnras{Mon.\ Not.\ R.\ Astron.\ Soc.\ }
\def\cqg{Classical\ Quantum\ Gravity\ }
\begin{document}
\title
{Determining the stiffness of the equation of state \break
using low $T/W$ dynamical instabilities in differentially rotating
stars}
%
\author{Motoyuki Saijo}
\email[E-mail: ]{saijo@aoni.waseda.jp}
%
\affiliation
{Research Institute for Science and Engineering, Waseda University,
  Shinjuku, Tokyo 169-8555, Japan}
%
\received{7 May 2018}
\accepted{18 June 2018}
%
\begin{abstract}
We investigate the nature of low $T/W$ dynamical instabilities in
various ranges of the stiffness of the equation of state in
differentially rotating stars.  Here $T$ is the rotational kinetic
energy, while $W$ the gravitational binding energy.  We analyze these
instabilities in both a linear perturbation analysis and a
three-dimensional hydrodynamical simulation.  An unstable normal mode
of a differentially rotating star is detected by solving an eigenvalue
problem along the equatorial plane of the star.  The physical
mechanism of low $T/W$ dynamical instabilities is also qualitatively
confirmed by a scattering of sound waves between corotation and the
surface caused by the corotation barrier.  Therefore, we can draw a
picture of existing pulsation modes unstabilized due to an amplified
reflection of sound waves from the corotation barrier. The feature in
the eigenfrequency and eigenfunction of the unstable mode in the
linear analysis roughly agrees with that in the three-dimensional
hydrodynamical simulation in Newtonian gravity.  Moreover, the nature
of the eigenfunction that oscillates between corotation and the
surface for an unstable star requires reinterpretation of pulsation
modes in differentially rotating stars.  Finally, we propose a manner
by which to constrain the stiffness of the equation of state by the
direct detection of mode decomposed gravitational waveforms.
\end{abstract}
\maketitle
%
\section{Introduction
\label{sec:intro}}
Low $T/W$ dynamical instabilities in differentially rotating stars
have been found numerically around the 21st century \citep{PDD96,
  CNLB01, SKE02, *SKE03, SBS03}.  Here $T$ is the rotational kinetic
energy, while $W$ is the gravitational binding energy.  In contrast to
the standard nonaxisymmetric rotational dynamical $m=2$ bar
instability (the threshold is $T/W$ = 0.2738 for a uniformly rotating,
incompressible star \citep{Chandra69, Tassoul78, ST83}, toroidal modes
$e^{im\varphi}$, where  $m=\pm1$, $\pm2$, $\ldots$, $\varphi$ is the
azimuthal angle), a star becomes dynamically unstable in a
significantly low magnitude of rotation when these instabilities set
in.  A star also becomes dynamically unstable to the $m=1$ spiral mode
\citep{CNLB01, SBS03}, which has never been found in a rotating
stellar configuration.  According to plenty of results from numerical
simulations, a moderate degree of a differentially rotating
configuration of the star is at least required to trigger these
instabilities, and the stiffness of the equation of state changes the
dominant behavior of these instabilities between spiral and bar
\citep{SBS03, SY06} (except for Ref.~\citep{OT06}).

There are two representative cases of astrophysical applications to
low $T/W$ dynamical instabilities.  One is the binary neutron star
mergers.  After the merger, a moderate degree of a differentially
rotating configuration can be generated in a dynamical timescale,
which may trigger these instabilities.  Recent numerical simulations
have found a spiral type of configuration after the merger (see, e.g.,
\citep{Paschalidis15, East16, Lehner16, Radice16}), and found the
angle direction changes in a constant phase curve
\citep{Paschalidis15, East16}, which may be essentially the same as
that in $m=1$ low $T/W$ dynamical instabilities \citep{SY06}.  The
other is the collapse driven supernovae.  After the core bounce, a
proto-neutron star with a high degree of differential rotation (see,
e.g., \citep{Dimmelmeier02a, *Dimmelmeier02b, KTK14}) can be generated
in a dynamical timescale.  In such a case, requirements for triggering
these instabilities are naturally satisfied.  In fact, the $m=1$
instabilities of gravitational waveforms in collapse driven supernovae
have been demonstrated (see, e.g., \citep{Ott07}).

In contrast to a clear existence of low $T/W$ dynamical instabilities,
a physical mechanism to trigger them is still a mystery.  The main
issue comes from the fact that there is no complete successful
multi-dimensional linear perturbation analysis in differentially
rotating stars.  At present, there are two representative discussions
about the necessary conditions to trigger these instabilities and
their demonstrations, mainly based on a self-gravitating disk system.
One is the corotation resonance, which originally comes from the
density wave theory that a self-gravitating disk becomes unstable due
to the absorption of angular momentum at corotation (see, e.g.,
\citep{LK72}).  Since corotation plays an essential role in a disk
system, a characteristic feature of corotation to these instabilities
in basic equations has been argued \citep{WAJ05}.  Comparison between
low $T/W$ dynamical instabilities and standard $m=2$ ones based on a
canonical angular momentum distribution has been discussed
\citep{SY06}.  The other is nonaxisymmetric Rossby wave instabilities,
which have been investigated for finding a necessary condition where a
potential vorticity takes an extreme \citep{LLCN99}.  The necessary
conditions acquired in a disk system have been applied to low $T/W$
dynamical instabilities \citep{OT06,Corvino10}.  However, no
multi-dimensional linear perturbation analysis has been done so far in
a rotating stellar configuration.  Such an analysis is necessary since
both a rotating stellar configuration and a self-gravitating effect
may take place (but see Refs.~\citep{SKE02, *SKE03, KE03, PA15} for a
specific fundamental mode of pulsating stars).  Without the analysis
mentioned above, it would not be possible to completely understand the
physical mechanism of low $T/W$ dynamical instabilities.

The purpose of this paper is threefold.  First, we want to understand
physical features of low $T/W$ dynamical instabilities by both linear
analyses and numerical simulations.  Although plenty of numerical
simulations have confirmed the existence of their instabilities,
understanding their features, such as extracting functional dependence
of characteristic frequencies and timescales, is extremely expensive
solely by numerical simulations.  At least, large parameter sets of
computations are required.  On the other hand, a full set of linear
perturbation analysis in differentially rotating stars requires two
spatial dimensional analyses even when we adopt harmonic expansion for
time and azimuthal angle.  It is still in progress in a general manner
even in Newtonian gravity (see, e.g., \citep{Unno89, KE03}).  But
instead of exploring these instabilities in a two-dimensional linear
perturbation analysis, we restrict our analysis to the equatorial
motion of a perturbed fluid, taking a self-gravitating effect into
account.  Although this is a crude assumption we impose in our study
\citep{SY16, YS17}, it would still be useful for finding some physical
aspects of these instabilities by combining two complementary
approaches.  This part is an extension work of Ref.~\citep{SY16} for a
wide set of parameters, each of which varies the stiffness of the
equation of state.

Second, we want to understand a physical mechanism of low $T/W$
dynamical instabilities.  Computational results from three-dimensional
hydrodynamical simulations are sufficiently attractive to understand
the dynamical features of their instabilities, but without
sophisticated diagnostic quantities and plenty of parameter searches,
it is extremely difficult to confirm the physical mechanisms by
themselves.  Although a linear analysis can only apply to the linear
stage of the instability growth, it is quite powerful to identify the
specific features of these instabilities.  Our idea is to investigate
the linear analysis in these systems and acquire a picture for
generating these instabilities.  Such a picture could be useful for a
deeper understanding of these instabilities by three-dimensional
numerical simulations.  Although the idea of a scattering problem by
the corotation barrier has been introduced in Ref.~\citep{YS17}, we
have improved and adjusted the analysis by comparing the results of
linear analyses with those of three-dimensional numerical
simulations.

Finally, we focus on gravitational waves generated from these
instabilities.  Nonaxisymmetric instabilities are subject to
quasi-periodic gravitational waves in general.  In principle, these
gravitational waves can be detected in ongoing ground-based detectors
such as advanced LIGO, advanced VIRGO, KAGRA, and future projects such
as Einstein Telescope \citep{Andersson13}.  All detectors have good
sensitivity around kilohertz frequencies, preparing for exploring the
dynamics of neutron stars.  In fact, recent detection of the merger of
binary neutron stars has opened a new era for exploring neutron stars
by gravitational waves \citep{gw170817a, *gw170817b}.  Detailed
analysis of gravitational waveforms may tell us a variety of interior
features of neutron stars, and it would be worth investigating
gravitational waveforms and their spectra of these instabilities from
a theoretical viewpoint.  Moreover, we propose a method to constrain
the stiffness of the equation of state from the direct observation of
gravitational waves, which would potentially become a guideline for
future realistic astrophysical simulations, direct observations, and
data analyses for extracting physics.

The content of this paper is as follows.  In Sec.~\ref{sec:hydro}, we
briefly explain the basic hydrodynamic equations in Newtonian gravity
equipping shock capturing schemes in our code with a code test.  In
Sec.~\ref{sec:perturbation}, we introduce our basic equations of
perturbative approaches and our formulation of a normal mode analysis
and a scattering problem rising from the corotation barrier, as well
as their results.  In Sec.~\ref{sec:results}, we show our results of
three-dimensional numerical simulations and compare them with those of
perturbative approaches and gravitational waves from low $T/W$
dynamical instabilities.  Section~\ref{sec:Conclusions} is devoted to
the summary of this paper.  Throughout this paper, we use the
geometrized units with $G=c=1$ \footnote{The speed of light only
  enters through the quadrupole formula of gravitational waves.}.

\section{Hydrodynamics
\label{sec:hydro}}
\subsection{Basic equations
\label{subsec:basic_hydro}}
We summarize here the basic equations for perfect fluids of
hydrodynamics in Newtonian gravity.  We assume an adiabatic
$\Gamma$-law equation of state
\begin{equation}
p = ( \Gamma - 1 ) \rho \varepsilon,
\label{eqn:GammaLaw}
\end{equation}
where $p$ is the pressure, $\Gamma$ the adiabatic index, $\rho$ the
rest mass density, and $\varepsilon$ the specific internal energy
density.  For perfect fluids, Newtonian equations of hydrodynamics
consist of the continuity equation
\begin{equation}
\frac{\partial \rho}{\partial t}
+\frac{\partial (\rho v^{j})}{\partial x^{j}} = 0,
\label{eqn:continuity}
\end{equation}
where $v^i$ is the velocity, the energy equation
\begin{equation}
\frac{\partial e}{\partial t}+
\frac{\partial [(e + p)v^{j}]}{\partial x^{j}} = 
- \rho v^j \frac{\partial \Phi}{\partial x^j} 
,
\end{equation}
where $e = \rho (\varepsilon + v_j v^j / 2)$ is the total energy, and
the Euler equations
\begin{equation}
\frac{\partial(\rho v^{i})}{\partial t}
+ \frac{\partial (\rho v^{i} v^{j} + p \delta^{ij} )}{\partial x^{j}} 
=
- \rho \frac{\partial \Phi}{\partial x^{i}},
\end{equation}
where $\Phi$ is the gravitational potential, which satisfies
\begin{equation}
\triangle \Phi = 4 \pi \rho.
\end{equation}

\begin{table}[!htb]
\begin{center}
\caption{
Equilibrium configuration of differentially rotating stars.}
\begin{ruledtabular}
\begin{tabular}{c c r c}
Model & $n$ & $\Omega_c / \Omega_e$\footnotemark[1]   & $T/W$
 \\
 \hline
I(a) & 1 & 26.0 & $6.09 \times 10^{-2}$\\
I(b) & 1 & 12.1 & $8.00 \times 10^{-2}$\\
I(c) & 1 & 5.0 & $1.00 \times 10^{-1}$\\
I(d) & 1 & 2.0 & $1.09 \times 10^{-1}$\\
II(a) & 1.5 & 26.0 & $6.76 \times 10^{-2}$\\
II(b) & 1.5 & 12.1 & $8.61 \times 10^{-2}$\\
II(c) & 1.5 & 5.0 & $1.01 \times 10^{-1}$\\
II(d) & 1.5 & 2.0 & $9.37 \times 10^{-2}$\\
III(a) & 2 & 26.0 & $7.29 \times 10^{-2}$\\
III(b) & 2 & 12.1 & $8.89 \times 10^{-2}$\\
III(c) & 2 & 5.0 & $9.38 \times 10^{-2}$\\
III(d) & 2 & 2.0 & $6.48 \times 10^{-2}$\\
IV(a) & 3 & 26.0 & $7.21 \times 10^{-2}$\\
IV(b) & 3 & 12.1 & $6.89 \times 10^{-2}$\\
IV(c) & 3 & 5.0 & $4.21 \times 10^{-2}$\\
IV(d) & 3 & 2.0 & $1.81 \times 10^{-2}$\\
\end{tabular}
\end{ruledtabular}
\label{tab:equilibrium}
\footnotetext[1]{$\Omega_c$: Central angular velocity; $\Omega_e$:
  Equatorial surface angular velocity.}
\end{center}
\end{table}

Constructing the equilibrium configuration of a differentially
rotating star assuming a polytropic equation of state $p = \kappa_{\rm
  s} \rho^{\Gamma}$ with $\kappa_{\rm s}$ being constant, first we
have to solve the Euler equations.  In the axisymmetric configuration,
the equation can be described in the cylindrical coordinates as
\begin{equation}
\frac{1}{\rho} \nabla_{\varpi} p + \nabla_{\varpi} \Phi - \varpi
\Omega^2 = 0,
\label{eqn:EulerCyl}
\end{equation}
where $\varpi$ is the cylindrical radius and $\Omega$ the angular
velocity.  Here we adopt $j$-constant rotation law for the angular
velocity distribution of the star as
\begin{equation*}
\Omega = \frac{j_0}{d^2 + \varpi^2},
\end{equation*}
where $j_0$ is the constant and $d$ the degree of differential
rotation.  With this rotation law, we can integrate
Eq.~(\ref{eqn:EulerCyl}) as
\begin{equation}
H + \Phi + \Phi_c = C,
\end{equation}
where the enthalpy $H$ and the rotational potential $\Phi_c$ are
\begin{eqnarray}
H &=& \varepsilon + \frac{p}{\rho},
\\
\Phi_c &=& - \int^{\varpi} \varpi \Omega^2~d\varpi
= \frac{1}{2} \frac{j_0^2}{d^2+\varpi^2}
,
\end{eqnarray}
$n$ is the polytropic index with a relation $\Gamma = 1 + 1/n$, and
$C$ the constant.  We summarize our configuration of differentially 
rotating stars in Table \ref{tab:equilibrium}.

\subsection{Shock capturing scheme and wall shock tests
\label{subsec:shock}}
Here we review the shock capturing scheme inserted in our Newtonian
hydrodynamics code.  The flux conservative form of the continuity
equation, the Euler equations, and the energy equation can be written
as
\begin{equation}
\frac{\partial}{\partial t} \boldsymbol{\cal U} +
\frac{\partial}{\partial x^j} \boldsymbol{\cal F}^j =
\boldsymbol{\cal S},
\end{equation}
where the state vector $\boldsymbol{\cal U}$, the flux vectors
$\boldsymbol{\cal F}^j$ and the source vector $\boldsymbol{\cal S}$
are
\begin{equation}
\boldsymbol{\cal U} = 
\left[
\begin{array}{c}
\rho \\
\rho v^i \\
e
\end{array}
\right], 
\,
\boldsymbol{\cal F}^j = 
\left[
\begin{array}{c}
\rho v^j \\
\rho v^i v^j + p \delta^{ij}\\
(e + p) v^j
\end{array}
\right], 
\,
\boldsymbol{\cal S} = 
\left[
\begin{array}{c}
0 \\
- \rho
  \frac{\displaystyle \partial \Phi}{\displaystyle \partial x^i} \\
- \rho v^j \frac{\displaystyle \partial \Phi}{\displaystyle \partial x^j}
\end{array}
\right].
\end{equation}

We use monotonized central-difference (MC) limiter \citep{LeVeque98}
for interpolating the conservative quantities on the grid to the
numerical cell boundaries.  To respect their thermodynamical
properties, we choose $\rho$, $v^i$, and $\varepsilon$ as primitive
quantities.  For given primitive variables $u_k$( $\equiv [\rho_k,
  v_k^i, \varepsilon_k]$, $k$: label of grid point), we are able to
interpolate the quantities to the left and right intercell boundaries
located at $k\pm1/2$.  We use the second-order accuracy with monotonic
piecewise linear slopes along the coordinate $x_k$ as
\begin{eqnarray*}
u_{k-1/2}^{R} &=& u_k + \sigma_k (x_{k-1/2} - x_k),
\\
u_{k+1/2}^{L} &=& u_k + \sigma_k (x_{k+1/2} - x_k),
\end{eqnarray*}
where 
\begin{equation}
\sigma_k = {\rm minmod} 
\left[
  2 \left( \frac{\Delta u_k }{\Delta x_k} \right), 
  2 \left( \frac{\Delta u_{k+1}}{\Delta x_{k+1}} \right)
\right],
\end{equation}
$\Delta u_k \equiv u_k - u_{k-1}$, $\Delta x_k \equiv x_k - x_{k-1}$
is the grid separation, and
\begin{eqnarray*}
&&{\rm minmod}[a,b]\\
&&\qquad =  \left\{
\begin{array}{ll}
0 & ab \leq 0,\\
{\rm sgn}(a) \min \left[ 
  2 |a|, 
  2 |b|, 
  \left( \frac{\displaystyle |a + b|}{\displaystyle 2} \right)
\right] & {\rm otherwise}.
\end{array}
\right.
\end{eqnarray*}
Note that the label $L$ and $R$, respectively, represent the left and
right sides of intercell boundaries located at $k\pm1/2$.

We adopt the approximate Harten-Lax-van Leer (HLL) Riemann solver
\citep{Toro09} for constructing a numerical flux
\begin{equation}
F_{\rm HLL}^j = \frac{S_R^j F_L^j - S_L^j F_R^j + S_L^j S_R^j (U_R^j -
  U_L^j)}{S_R^j - S_L^j},
\end{equation}
where $F_{L,R}^j$ are the flux vectors at the left and right numerical
cells, and $S_{L,R}^j$ are the characteristic speeds at the left and
right intercell boundaries determined as
\begin{eqnarray}
S_L^j &=& \max (0, \lambda^{j+}_{L}, \lambda^{j+}_{R}, v^{j}_{L},
v^{j}_{R})
,\\
S_R^j &=& \min (0, \lambda^{j-}_{L}, \lambda^{j-}_{R}, v^{j}_{L},
v^{j}_{R})
.
\end{eqnarray}
The quantities $\lambda^{j\pm}$ are the maximum and the minimum of the
eigenvalues in the Jacobian matrix of the flux vectors as
\begin{equation}
\lambda^{j\pm} = v^j \pm c_s,
\end{equation}
where $c_s$ is a speed of sound.

\begin{figure}
\centering
\includegraphics[keepaspectratio=true,width=8cm]{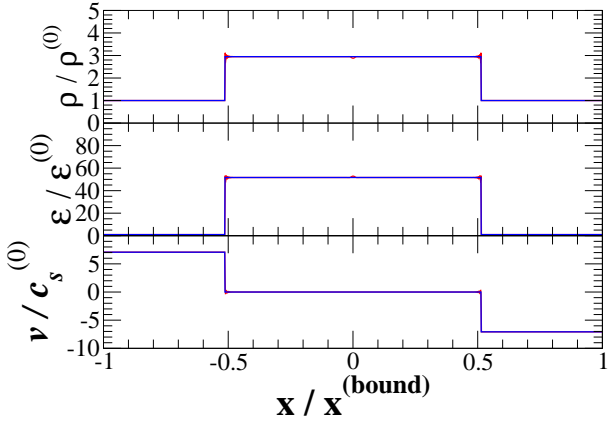}
\caption{
Comparison of a rest mass density, specific internal energy and
velocity between numerical and analytical results of the
one-dimensional wall shock problem at $t = 1.0~x^{\rm(bound)} / v_0$.
Red and blue lines represent our computational and analytical results.
We choose the parameter sets as $\Gamma = 2$, $\kappa_{\rm s} = 1$
with grid space $\Delta x = 1.0 \times 10^{-3} x^{\rm(bound)}$ and
$v_0= 7.07~c_s^{(0)}$, where $c_s^{(0)}$ is the initial speed of sound
at $t=0$.
\label{fig:wshock}
}
\end{figure}

We check the ability of our new HLL flux scheme to resolve shocks by
solving a wall shock problem, in which two phases of a fluid collide
at supersonic speeds.  In Fig.~\ref{fig:wshock}, we compare numerical
results with the analytic solutions for initial velocities that are
similar to those found in our simulations in
Sec.~\ref{sec:results}.  We find good agreement for Mach numbers up to 
$M_{\rm mach} \lesssim 7$, which is a typical number in our
simulations.

\section{Linear Perturbation
\label{sec:perturbation}}
\subsection{Basic equations in a nonaxisymmetric perturbation
\label{subsec:bequation}}
We perturb the differentially rotating stars nonaxisymmetrically in
order to investigate the feature of low $T/W$ dynamical
instabilities.  We assume a harmonic dependence of time and azimuthal
angle on the perturbed quantity $\delta q$ as
\begin{equation}
\delta q(t,\varpi,z,\varphi) = \sum_m \delta q_m (\varpi, z)
e^{-i\omega t + i m \varphi},
\end{equation}
where $z$ is the coordinate along the rotational axis, $\varphi$ the
azimuthal coordinate, and $m$ the azimuthal wave number.  The
perturbed Euler equations can be written as \citep{IL89}
\begin{equation}
Q_{ij}^{-1} \delta v^{j} \equiv [i \tilde{\omega} \gamma_{ij} - 2
  \nabla_{j} v_{i} + \phi_{i} \nabla_{j} \Omega ] \delta
v^j = \nabla_i \delta U,
\label{eqn:PEuler}
\end{equation}
where $\delta v^i$ is the perturbed velocity, $v^i$ the equilibrium
distribution of the velocity, $\tilde{\omega}=\omega - m \Omega$,
$\gamma_{ij}$ the spatial 3-metric, $\phi^i$ the rotational Killing
vector, $\delta U$ the scalar potential $\delta U \equiv \delta h +
\delta \Phi$, $\delta h$ the perturbed enthalpy, and $\delta \Phi$ the
perturbed gravitational potential.  Note that we define the tensorial
quantity $Q_{ij}$ in Eq.~(\ref{eqn:PEuler}).

Assuming a barotropic relation $p=p(\rho)$ in the equation of state, a
perturbed rest mass density $\delta \rho$ can be written as
\begin{equation}
\delta \rho = \rho \frac{d\rho}{dp} \delta h = \rho \frac{d\rho}{dp}
(\delta U - \delta \Phi).
\label{eqn:drho}
\end{equation}
Using Eq.~(\ref{eqn:drho}), the conservation equation of the perturbed
rest mass is described using $\delta U$ and $\delta \Phi$ as
\begin{equation}
- i\tilde{\omega} \rho \frac{d\rho}{dp} (\delta U - \delta \Phi) +
\nabla_{i} (\rho Q^{ij} \nabla_j \delta U) = 0.
\label{eqn:PContinuity}
\end{equation}
Combining Eqs.~(\ref{eqn:PEuler}) and (\ref{eqn:PContinuity}), one of
the pulsation equations of rotating stars becomes second-order
partially differential equations as
\begin{eqnarray}
&&\left[
\frac{\partial^2}{\partial\varpi^2} - 
\left( \frac{\partial}{\partial\varpi} \ln \frac{D}{\rho \varpi}
\right) \frac{\partial}{\partial\varpi}
 - \frac{2m\Omega}{\varpi \tilde{\omega}}
\left( \frac{\partial}{\partial\varpi} \ln \frac{\rho \Omega}{D}
\right) - \frac{m^2}{\varpi^2} 
\right. \nonumber \\
&&
\hspace{1cm}
\left.
- \frac{D}{\tilde{\omega}^2}
  \left( \frac{\partial^2}{\partial z^2} 
    + \frac{1}{\rho} \frac{\partial \rho}{\partial z}
    \frac{\partial}{\partial z} \right)
- \frac{D}{dp/d\rho}
\right]
\delta U_m (\varpi, z) 
\nonumber \\
&&
\hspace{1cm}
= - \frac{D}{dp/d\rho} \delta\Phi_m (\varpi, z)
\label{eqn:Basic_dU}
,
\end{eqnarray}
where $D = \kappa^2 - \tilde{\omega}^2$ and  $\kappa^2$ is $\varpi
(d\Omega^2 / d\varpi) + 4 \Omega^2$.  The perturbed Poisson's equation
is
\begin{equation}
\nabla_{j} \nabla^{j} \delta \Phi = 4 \pi \rho \frac{d\rho}{dp}
(\delta U - \delta \Phi),
\end{equation}
and it is explicitly expressed as 
\begin{eqnarray}
&&\left[
\frac{\partial^2}{\partial\varpi^2} + 
\frac{1}{\varpi} \frac{\partial}{\partial\varpi}
- \frac{m^2}{\varpi^2} + \frac{\partial^2}{\partial z^2}
+ 4 \pi \rho \frac{d\rho}{dp} 
\right]
\delta \Phi_m (\varpi, z) 
\nonumber \\
&&
\hspace{1cm}
= 4\pi\rho \frac{d\rho}{dp} \delta U_m (\varpi, z).
\label{eqn:Basic_dPhi}
\end{eqnarray}
To conclude, the basic pulsation equations of $\delta U_m$ and $\delta
\Phi_m$ are Eqs.~(\ref{eqn:Basic_dU}) and (\ref{eqn:Basic_dPhi}).

\subsection{Cylindrical model
\label{subsec:cylinder}}
We impose one assumption in which the equatorial motion of the
perturbed quantities of the stars alone is taken into account.  Our
basic idea is that a characteristic wave propagation mainly lies in
the equatorial plane in a rotating configuration.  Therefore, we
simply discard the second-order $z$ derivatives in $\delta U_m$ and
$\delta \Phi_m$ (the first-order $z$ derivatives in $\delta U_m$ and
$\delta \Phi_m$ automatically disappear due to an equatorial symmetry
which we imposed in the system).  We call this system a cylindrical
model.

\begin{figure*}
\centering
\includegraphics[keepaspectratio=true,width=16cm]{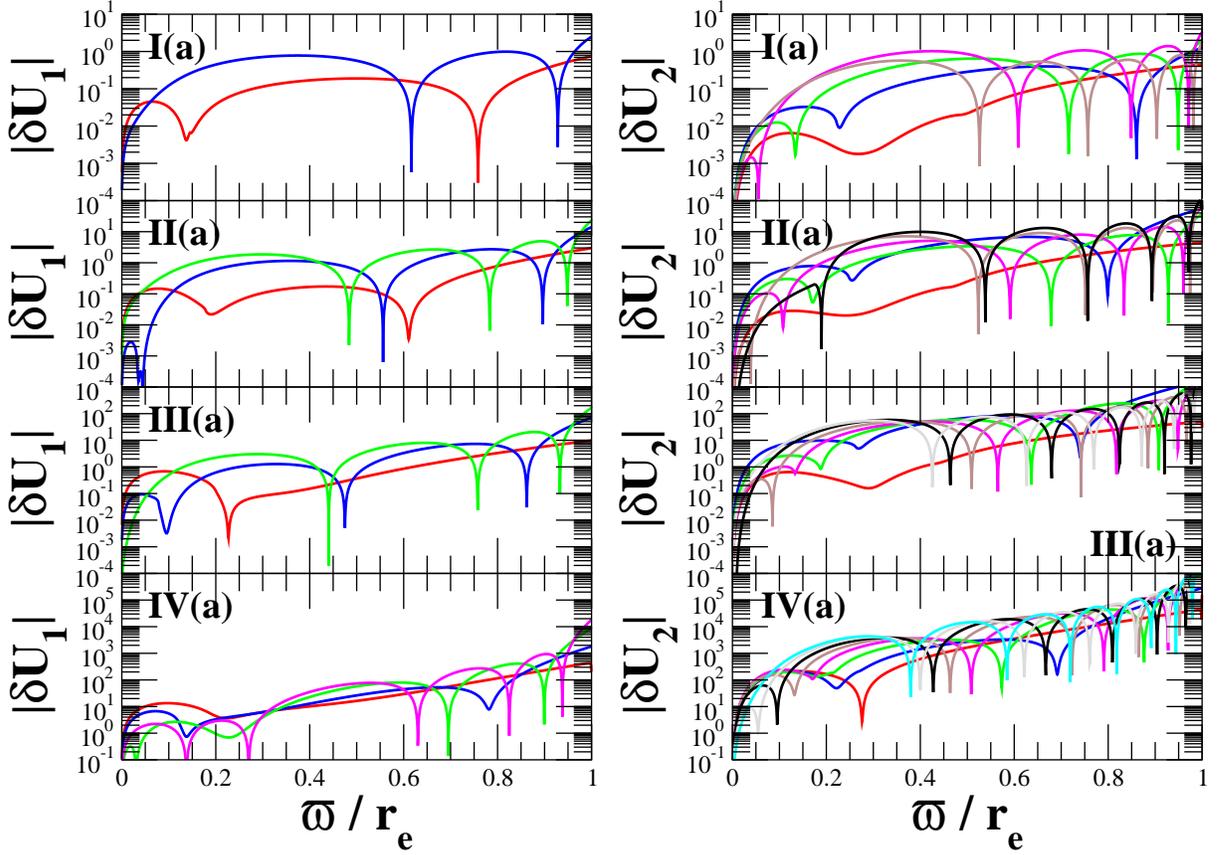}
\caption{
The $m=1$ and $m=2$ eigenfunctions $|\delta U_m|$ for four low $T/W$
dynamically unstable stars in cylindrical models.  The labels I(a),
II(a), III(a), and IV(a), respectively, represent the equilibrium
models in Table \ref{tab:equilibrium}.  Red, blue, green, magenta,
brown, black, grey, and cyan, respectively, represent the node number
between corotation and surface of $N=0$, $1$, $2$, $3$, $4$, $5$, $6$,
and $7$.  The increasing number of nodes between corotation and the
surface can clearly be seen in all eigenfunctions.
\label{fig:efd02p150cy}
}
\end{figure*}

The basic equations [Eqs.~(\ref{eqn:Basic_dU}) and
  (\ref{eqn:Basic_dPhi})] in the cylindrical model become (see, e.g.,
\citep{TL08, SY16})
\begin{eqnarray}
&&\left[
\frac{d^2}{d\varpi^2} - 
\left( \frac{d}{d\varpi} \ln \frac{D}{\rho \varpi} \right)
\frac{d}{d\varpi}
 - \frac{2m\Omega}{\varpi \tilde{\omega}}
\left( \frac{d}{d\varpi} \ln \frac{\rho \Omega}{D} \right) 
- \frac{m^2}{\varpi^2} 
\right. \nonumber \\
&&
\hspace{1cm}
\left.
- \frac{D}{dp/d\rho}
\right]
\delta U_m (\varpi) 
= - \frac{D}{dp/d\rho} \delta\Phi_m (\varpi)
\label{eqn:Basic_dU_cy}
,\\
&&\left[
\frac{d^2}{d\varpi^2} + 
\frac{1}{\varpi} \frac{d}{d\varpi}
 - \frac{m^2}{\varpi^2} + 4 \pi \rho \frac{d\rho}{dp} 
\right]
\delta \Phi_m (\varpi) 
\nonumber \\
&&
\hspace{1cm}
= 4\pi\rho \frac{d\rho}{dp} \delta U_m (\varpi).
\label{eqn:Basic_dPhi_cy}
\end{eqnarray}

We introduce an eigenvalue problem for studying the stability of the
system.  We impose regularity conditions at the center for $\delta
U_m$ and $\delta \Phi_m$ as
\begin{equation}
\delta U_m = C_m^1 \varpi^{|m|},
\hspace{5mm}
\delta \Phi_m = C_m^2 \varpi^{|m|},
\label{eqn:cyl_bdcentre}
\end{equation}
where $C_m^1$ and $C_m^2$ are constants.  We also impose a boundary
condition for a perturbed gravitational potential $\delta \Phi_m$ at
infinity as the quantity is finite ($\delta \Phi_m \propto
\varpi^{-|m|}$).  That is to say, we can equivalently impose a
boundary condition on the surface as
\begin{equation}
\delta \Phi_m = C_m^3 \varpi^{-|m|},
\label{eqn:cyl_bdsurfaceP}
\end{equation}
in our model.  Using the nature of linear perturbation, the constant
$C_{m}^{3}$ is described as an appropriate combination of $C_m^1$ and
$C_m^2$, which is determined from the condition that $\delta \Phi_m$
and $d \delta \Phi_m / d\varpi$ are continuous across the surface.  In
practice, we can construct two sets of solutions ($\delta U_m^{(1)}$,
$\delta \Phi_m^{(1)}$) and ($\delta U_m^{(2)}$, $\delta \Phi_m^{(2)}$)
by integrating Eqs. (\ref{eqn:Basic_dU_cy}) and
(\ref{eqn:Basic_dPhi_cy}), keeping the same $C_m^1$ but different
$C_m^2$ ($C_m^{2(1)}$ and $C_m^{2(2)}$) from the center to the
surface.  Although these solutions do not satisfy the boundary
condition on the surface [Eq.~(\ref{eqn:cyl_bdsurfaceP})] in general,
we are able to construct a solution by linearly combining these two
sets of solutions as
\begin{eqnarray}
  \delta U_m &=&
    p_m^{(1)} \delta U_m^{(1)} + p_m^{(2)} \delta U_m^{(2)}, \\
  \delta \Phi_m &=&
    p_m^{(1)} \delta \Phi_m^{(1)} + p_m^{(2)} \delta \Phi_m^{(2)},
\end{eqnarray}
where $p_m^{(1)}$ and $p_m^{(2)}$ should satisfy
\begin{widetext}
\begin{equation}
p_m^{(1)} \left(
  \left. \delta \Phi_m^{(1)} \right|_{\varpi = r_e} 
  + \frac{r_e}{m} \left. \frac{d\Phi_m^{(1)}}{d\varpi} \right|_{\varpi = r_e}
  \right) 
+ p_m^{(2)} \left(
  \left. \delta \Phi_m^{(2)} \right|_{\varpi = r_e} 
  + \frac{r_e}{m}\left. \frac{d\Phi_m^{(2)}}{d\varpi}\right|_{\varpi = r_e}
  \right) = 0,
\end{equation}
\end{widetext}
to meet the surface boundary condition for $\delta \Phi_m$
[Eq.~(\ref{eqn:cyl_bdsurfaceP})].  Note that $r_e$ represents the
equatorial radius of the star.  The constant $C_m^3$ can then be
written by using $p_m^{(1)}$ and $p_m^{(2)}$ as
\begin{equation}
C_m^3 = r_e^{|m|} 
\left(
p_m^{(1)} \left.\delta \Phi_m^{(1)} \right|_{\varpi = r_e} 
+ p_m^{(2)} \left. \delta \Phi_m^{(2)} \right|_{\varpi = r_e}
\right),
\end{equation}
remaining as one scaling freedom.

\begin{figure}
\centering
\includegraphics[keepaspectratio=true,width=8cm]{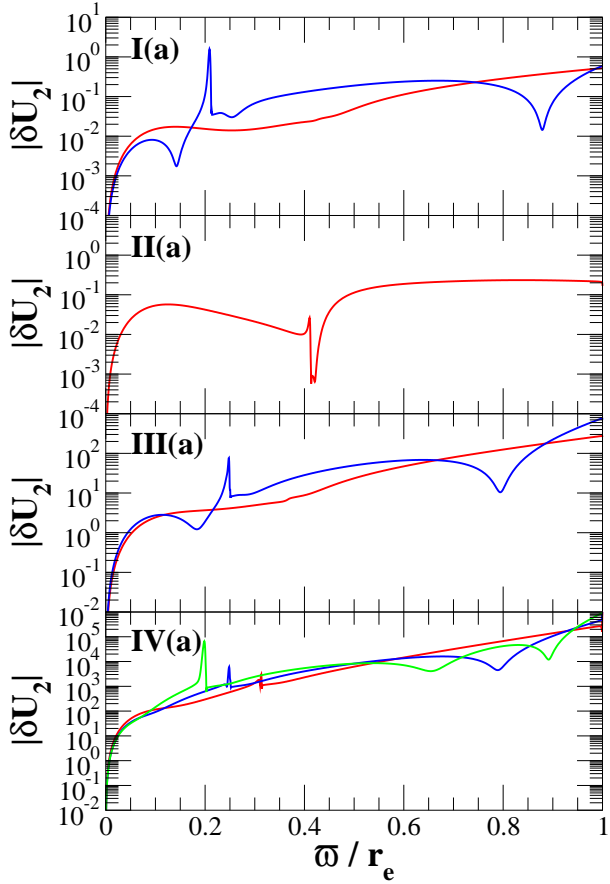}
\caption{
Same as Fig.~\ref{fig:efd02p150cy}, but for $m=2$ eigenfunctions
$|\delta U_m|$ in spheroidal models.
\label{fig:efd02p150sp}
}
\end{figure}

We also impose a surface boundary condition for $\delta U_m$ as the
enthalpy vanishes on the oscillating surface.  Namely
\begin{equation}
\delta h_m + \xi_m^{j} \nabla_{j} h = 0,
\label{eqn:bc_cy}
\end{equation}
where $h$ is the equilibrium enthalpy, $\xi_m^i$ the Lagrangian
displacement \citep{IL90} of the $m$ mode as
\begin{equation}
\xi_{mi} = i \left[ \frac{\gamma_{ij}}{\tilde{\omega}} + i
  \frac{\phi_i \nabla_j \Omega}{\tilde{\omega}^2}\right] \delta v^j_m
,
\end{equation}
and $\delta v^i_m$ is the perturbed velocity of the $m$ mode.  A
concrete boundary condition for $\delta U_m$ along the equatorial
plane in the cylindrical coordinates is written as
\begin{table}[!htb]
\begin{center}
\caption{
The $m=1$ and $m=2$ normal modes of $n=1$ differentially rotating
stars in the cylindrical model.}
\begin{ruledtabular}
\begin{tabular}{c c c c c c c}
Model & $m$ & $N$\footnotemark[1]  & $\Re [\omega] / \Omega_c $ & $\Im 
[\omega] / \Omega_c $ &
$r_{\rm cr} / r_e$\footnotemark[2] &
$\Im[\omega]_{\rm ref} / \Omega_c$\footnotemark[3] 
\\
\hline
I(a) & $1$ & 1 & 0.65275 & 0.00012 & 0.14587 & 0.00001\\
I(a) & $1$ & 2 & 1.15197 & 0.00000 & $\cdots$ & $\cdots$\\
\hline
I(a) & $2$ & 0 & 0.29017 & 0.00855 & 0.48549 & 0.00110\\
I(a) & $2$ & 1 & 0.89791 & 0.00100 & 0.22158 & 0.00005\\
I(a) & $2$ & 2 & 1.39597 & 0.00039 & 0.13156 & 0.00001\\
I(a) & $2$ & 3 & 1.86074 & 0.00024 & 0.05471 & 0.00000\\
I(a) & $2$ & 4 & 2.31080 & 0.00000 & $\cdots$ & $\cdots$\\
\hline
I(b) & $1$ & 1 & 0.53732 & 0.00001 & 0.27838 & 0.00000\\
I(b) & $1$ & 2 & 1.77131 & 0.00000 & $\cdots$ & $\cdots$\\
\hline
I(b) & $2$ & 0 & 0.42797 & 0.01568 & 0.57497 & 0.00122\\
I(b) & $2$ & 1 & 1.40812 & 0.00036 & 0.19450 & 0.00001\\
I(b) & $2$ & 2 & 2.17266 & 0.00000 & $\cdots$ & $\cdots$\\
\hline
I(c) & $1$ & 0 & 1.20707 & 0.00000 & $\cdots$ & $\cdots$\\
\hline
I(c) & $2$ & 0 & 0.56668 & 0.00001 & 0.79519 & NA\footnotemark[4] \\
I(c) & $2$ & 1 & 2.31692 & 0.00000 & $\cdots$ & $\cdots$\\
\hline
I(d) & $1$ & 0 & 2.23106 & 0.00000 & $\cdots$ & $\cdots$\\
\hline
I(d) & $2$ & 0 & 0.70460 & 0.22153 & $\cdots$ & $\cdots$\\
I(d) & $2$ & 1 & 3.66914 & 0.00000 & $\cdots$ & $\cdots$\\
\end{tabular}
\end{ruledtabular}
\label{tab:qnm_cyp150-1}
\footnotetext[1]{$N$: Node numbers between corotation and equatorial
  surface radius.}
\footnotetext[2]{$r_{\rm cr}$: Corotation radius.}
\footnotetext[3]{$\Im[\omega]_{\rm ref}$: Imaginary part of the complex 
  frequency computed from the amplification timescale.}
\footnotetext[4]{NA: No amplification.}
\end{center}
\end{table}
\begin{eqnarray}
  &&\delta U_m - \delta \Phi_m
  - \frac{1}{D} (\nabla_{\varpi} \delta U_m )
  (\nabla_{\varpi} U - \nabla_{\varpi} \Phi )  
\nonumber \\
&&
\hspace{1cm}
+ \frac{2m\Omega}{\varpi \tilde{\omega} D} 
(\nabla_{\varpi} U - \nabla_{\varpi} \Phi) \delta U_m = 0.
\label{eqn:BCconcrete_cy}
\end{eqnarray}
Note that we impose a planner symmetry across the equatorial plane for
an equilibrium configuration to derive Eq.~(\ref{eqn:BCconcrete_cy}).
We also apply the Euler equations [Eq.~(\ref{eqn:EulerCyl})] to
compute the term $\nabla_{\varpi} U - \nabla_{\varpi} \Phi$ on the
boundary as
\begin{equation}
\nabla_{\varpi} U - \nabla_{\varpi} \Phi = - \nabla_{\varpi} \Phi +
\varpi \Omega^2.
\end{equation}
Only 1 degree of freedom remains in the system, which represents the
normalization factor in linear perturbation.  We set $C_m^1=1$ in our
computational code, closing the system as an eigenvalue problem.

\begin{table}[!htb]
\begin{center}
\caption{
Same as Table~\ref{tab:qnm_cyp150-1}, but of $n=1.5$ differentially
rotating stars.}
\begin{ruledtabular}
\begin{tabular}{c c c c c c c}
Model & $m$ & $N$ & $\Re [\omega] / \Omega_c $ & $\Im [\omega] /
\Omega_c $ & $r_{\rm cr} / r_e$ & $\Im[\omega]_{\rm ref} / \Omega_c$
\\
\hline
II(a) & $1$ & 1 & 0.55723 & 0.00080 & 0.17828 & 0.00016\\
II(a) & $1$ & 2 & 0.97454 & 0.00019 & 0.03233 & 0.00000\\
II(a) & $1$ & 3 & 1.34786 & 0.00000 & $\cdots$ & $\cdots$\\
\hline
II(a) & $2$ & 0 & 0.31734 & 0.00841 & 0.46054 & 0.00001\\
II(a) & $2$ & 1 & 0.78994 & 0.00196 & 0.24753 & 0.00007\\
II(a) & $2$ & 2 & 1.18978 & 0.00026 & 0.16504 & 0.00037\\
II(a) & $2$ & 3 & 1.56573 & 0.00044 & 0.10533 & 0.00008\\
II(a) & $2$ & 4 & 1.92897 & 0.00001 & 0.03838 & 0.00000\\
II(a) & $2$ & 5 & 2.00084 & 0.00000 & $\cdots$ & $\cdots$\\
\hline
II(b) & $1$ & 1 & 0.53737 & 0.00001 & 0.27836 & NA\\
II(b) & $1$ & 2 & 1.48459 & 0.00000 & $\cdots$ & $\cdots$\\
\hline
II(b) & $2$ & 0 & 0.44482 & 0.00723 & 0.56094 & 0.00003\\
II(b) & $2$ & 1 & 1.22913 & 0.00024 & 0.23758 & 0.00147\\
II(b) & $2$ & 2 & 1.83866 & 0.00042 & 0.08887 & 0.00001\\
II(b) & $2$ & 3 & 2.00188 & 0.00000 & $\cdots$ & $\cdots$\\
\hline
II(c) & $1$ & 0 & 1.20707 & 0.00000 & $\cdots$ & $\cdots$\\
\hline
II(c) & $2$ & 0 & 0.58081 & 0.00001 & 0.78158 & 0.00332\\
II(c) & $2$ & 1 & 1.70706 & 0.00001 & 0.20713 & 0.00031\\
II(c) & $2$ & 2 & 2.00116 & 0.00000 & $\cdots$ & $\cdots$\\
\hline
II(d) & $1$ & 0 & 2.17560 & 0.00000 & $\cdots$ & $\cdots$\\
\hline
II(d) & $2$ & 0 & 0.31532 & 0.00000 & $\cdots$ & $\cdots$\\
II(d) & $2$ & 1 & 1.17702 & 0.00001 & 0.83619 & NA\\
II(d) & $2$ & 2 & 3.03110 & 0.00000 & $\cdots$ & $\cdots$\\
\end{tabular}
\end{ruledtabular}
\label{tab:qnm_cyp150-2}
\end{center}
\end{table}

\begin{table}[!htb]
\begin{center}
\caption{
Same as Table~\ref{tab:qnm_cyp150-1}, but of $n=2$ differentially
rotating stars.}
\begin{ruledtabular}
\begin{tabular}{c c c c c c c}
Model & $m$ & $N$ & $\Re [\omega] / \Omega_c $ & $\Im
[\omega] / \Omega_c $ &
$r_{\rm cr} / r_e$ &
$\Im[\omega]_{\rm ref} / \Omega_c$
\\
\hline
III(a) & $1$ & 1 & 0.50040 & 0.00553 & 0.19984 & 0.00053\\
III(a) & $1$ & 2 & 0.86641 & 0.00018 & 0.07853 & 0.00001\\
III(a) & $1$ & 3 & 1.18793 & 0.00000 & $\cdots$ & $\cdots$\\
\hline
III(a) & $2$ & 0 & 0.35218 & 0.00676 & 0.43262 & 0.00012\\
III(a) & $2$ & 1 & 0.73570 & 0.00249 & 0.26218 & 0.00056\\
III(a) & $2$ & 2 & 1.06871 & 0.00027 & 0.18670 & NA\\
III(a) & $2$ & 3 & 1.38662 & 0.00001 & 0.13302 & 0.00001\\
III(a) & $2$ & 4 & 1.69457 & 0.00001 & 0.08491 & 0.00000\\
III(a) & $2$ & 5 & 1.99668 & 0.00001 & 0.00816& NA\\
III(a) & $2$ & 6 & 2.29444 & 0.00000 & $\cdots$ & $\cdots$\\
\hline
III(b) & $1$ & 1 & 0.74278 & 0.00635 & 0.17654 & 0.00015\\
III(b) & $1$ & 3 & 1.30342 & 0.00000 & $\cdots$ & $\cdots$\\
\hline
III(b) & $2$ & 0 & 0.47957 & 0.00006 & 0.53417 & 0.00639\\
III(b) & $2$ & 1 & 1.13383 & 0.00024 & 0.26221 & 0.00002\\
III(b) & $2$ & 2 & 1.63501 & 0.00040 & 0.14174 & 0.00000\\
III(b) & $2$ & 3 & 2.10710 & 0.00000 & $\cdots$ & $\cdots$\\
\hline
III(c) & $1$ & 0 & 1.20716 & 0.00000 & $\cdots$ & $\cdots$\\
\hline
III(c) & $2$ & 0 & 0.81573 & 0.00002 & 0.60245 & NA\\
III(c) & $2$ & 1 & 1.82914 & 0.00027 & 0.15282 & 0.00002\\
III(c) & $2$ & 2 & 2.56178 & 0.00000 & $\cdots$ & $\cdots$\\
\hline
III(d) & $1$ & 0 & 2.14064 & 0.00000 & $\cdots$ & $\cdots$\\
\hline
III(d) & $2$ & 0 & 0.16891 & 0.00000 & $\cdots$ & $\cdots$\\
III(d) & $2$ & 1 & 2.23895 & 0.00000 & $\cdots$ & $\cdots$\\
\end{tabular}
\end{ruledtabular}
\label{tab:qnm_cyp150-3}
\end{center}
\end{table}

\subsection{Spheroidal model
\label{subsec:spherical}}
Here we introduce a spheroidal model to take a stellar configuration
partially into account.  Instead of discarding $z$ derivative in the
basic equation of $\delta U_m$ in the cylindrical model, we assume a
solution for the polar direction $\theta$ in the spherical coordinates
as Legendre polynomial $P_{l}^{m}(\cos\theta)$,
\begin{eqnarray}
\delta U &=& \sum_{l,m} \delta U_{lm}(r) P_{l}^{m}(\cos\theta)
e^{-i\omega t + i m \varphi}
,\\
\delta \Phi &=& \sum_{l,m} \delta \Phi_{lm}(r) P_{l}^{m}(\cos\theta)
e^{-i\omega t + i m \varphi}.
\end{eqnarray}
The basic equations in the spheroidal model are written as (see, e.g.,
\citep{IL90})
\begin{eqnarray}
&&
\left[
  \frac{d^2}{d r^2} + 
  \left[
    \frac{1}{r} \left(2 - \frac{\kappa^2}{\tilde{\omega}^2} \right) +
    \frac{1}{\rho} \frac{d \rho}{d r} -
    \frac{1}{D}\frac{\partial D}{d \varpi}
  \right] \frac{d}{d r}
\right.
\nonumber \\
&&
\hspace{1cm}
  - \frac{2m\Omega}{ r \tilde{\omega}}
\left( 
  \frac{1}{\rho} \frac{d \rho}{d r} 
  + \frac{1}{\Omega}\frac{d \Omega}{d \varpi}
   - \frac{1}{D}\frac{d D}{d \varpi}
\right) 
- \frac{D}{dp/d\rho}
\nonumber \\
&&
\hspace{1cm}
\left.
- \frac{1}{ r^2} 
\left[
  m^2 + \left( 1 - \frac{\kappa^2}{\tilde{\omega}^2} \right)
  [l (l+1) - m^2]
\right]
\right]
\delta U_{lm} (r) 
\nonumber \\
&&
\hspace{1cm}
= - \frac{D}{dp/d\rho} \delta\Phi_{lm} (r)
\label{eqn:Basic_dU_sp}
,\\
&&\left[
\frac{d^2}{d r^2} + 
\frac{2}{ r} \frac{d}{d r}
 - \frac{l (l+1)}{ r^2} + 4 \pi \rho \frac{d\rho}{dp} 
\right]
\delta \Phi_{lm} (r) 
\nonumber \\
&&
\hspace{1cm}
= 4\pi\rho \frac{d\rho}{dp} \delta U_{lm} (r).
\label{eqn:Basic_dPhi_sp}
\end{eqnarray}

\begin{table}[!thb]
\begin{center}
\caption{
Same as Table~\ref{tab:qnm_cyp150-1}, but of $n=3$ differentially
rotating stars.}
\begin{ruledtabular}
\begin{tabular}{c c c c c c c}
Model & $m$ & $N$ & $\Re [\omega] / \Omega_c $ & $\Im
[\omega] / \Omega_c $ &
$r_{\rm cr} / r_e$ &
$\Im[\omega]_{\rm ref} / \Omega_c$
\\
\hline
IV(a) & $1$ & 0 & 0.52603 & 0.05126 & 0.18985 & 0.00090\\
IV(a) & $1$ & 1 & 0.72759 & 0.00688 & 0.12238 & 0.00013\\
IV(a) & $1$ & 2 & 0.97965 & 0.00011 & 0.02883 & 0.00000\\
IV(a) & $1$ & 3 & 1.24636 & 0.00000 & $\cdots$ & $\cdots$\\
\hline
IV(a) & $2$ & 0 & 0.52656 & 0.00041 & 0.33456 & 0.00014\\
IV(a) & $2$ & 1 & 0.71208 & 0.00147 & 0.26897 & 0.00007\\
IV(a) & $2$ & 2 & 0.94196 & 0.00072 & 0.21197 & 0.00000\\
IV(a) & $2$ & 3 & 1.17403 & 0.00039 & 0.16775 & 0.00001\\
IV(a) & $2$ & 4 & 1.40432 & 0.00037 & 0.13026 & 0.00000\\
IV(a) & $2$ & 5 & 1.63112 & 0.00038 & 0.09511 & NA\\
IV(a) & $2$ & 6 & 1.85820 & 0.00046 & 0.05525 & 0.00000\\
IV(a) & $2$ & 7 & 2.08739 & 0.00000 & $\cdots$ & $\cdots$\\
\hline
IV(b) & $1$ & 0 & 0.80474 & 0.00005 & 0.14777 & 0.00058\\
IV(b) & $1$ & 1 & 1.10578 & 0.00000 & $\cdots$ & $\cdots$\\
\hline
IV(b) & $2$ & 0 & 0.80198 & 0.00001 & 0.36667 & 0.00018\\
IV(b) & $2$ & 1 & 1.10851 & 0.00013 & 0.26904 & 0.00019\\
IV(b) & $2$ & 2 & 1.42299 & 0.00026 & 0.19103 & 0.00001\\
IV(b) & $2$ & 3 & 1.74841 & 0.00051 & 0.11380 & NA\\
IV(b) & $2$ & 4 & 2.07242 & 0.00000 & $\cdots$ & $\cdots$\\
\hline
IV(c) & $1$ & 1 & 1.20713 & 0.00000 & $\cdots$ & $\cdots$\\
\hline
IV(c) & $2$ & 1 & 1.70729 & 0.00001 & 0.20703 & 0.00000\\
IV(c) & $2$ & 2 & 2.30367 & 0.00000 & $\cdots$ & $\cdots$\\
\hline
IV(d) & $1$ & 1 & 1.47885 & 0.00000 & $\cdots$ & $\cdots$\\
\hline
IV(d) & $2$ & 1 & 0.16891 & 0.00000 &  $\cdots$ & $\cdots$\\
IV(d) & $2$ & 2 & 3.56353 & 0.00000 &  $\cdots$ & $\cdots$\\
\end{tabular}
\end{ruledtabular}
\label{tab:qnm_cyp150-4}
\end{center}
\end{table}

\begin{table}[!htb]
\begin{center}
\caption{
The $m=2$ normal modes of differentially rotating stars in the
spheroidal model.}
\begin{ruledtabular}
\begin{tabular}{c c c c c c c}
Model & $n$ & $m$ & $N$ & $\Re [\omega] / \Omega_c $ & $\Im
[\omega] / \Omega_c $ &
$r_{\rm cr} / r_e$
\\
\hline
I(a) & $1$ & $2$ & 0 & 0.35170 & 0.01684 & 0.43297\\
I(a) & $1$ & $2$ & 1 & 0.93223 & 0.02054 & 0.21405\\
I(b) & $1$ & $2$ & 0 & 0.54951 & 0.00635 & 0.48741\\
I(c) & $1$ & $2$ & 0 & 0.91189 & 0.06664 & 0.54618\\
I(d) & $1$ & $2$ & 0 & $\cdots$ & $\cdots$ & $\cdots$\\
\hline
II(a) & $1.5$ & $2$ & 0 & 0.37904 & 0.00381 & 0.41359\\
II(b) & $1.5$ & $2$ & 0 & 0.53495 & 0.00586 & 0.49516\\
II(c) & $1.5$ & $2$ & 0 & $\cdots$ &$\cdots$ & $\cdots$\\
II(d) & $1.5$ & $2$ & 0 & $\cdots$ &$\cdots$ & $\cdots$\\
\hline
III(a) & $2$ & $2$ & 0 & 0.44143 & 0.01844 & 0.37580\\
III(a) & $2$ & $2$ & 1 & 0.77193 & 0.01448 & 0.25199\\
III(b) & $2$ & $2$ & 0 & 0.68370 & 0.03313 & 0.41626\\
III(c) & $2$ & $2$ & 0 & $\cdots$ & $\cdots$ & $\cdots$\\
III(d) & $2$ & $2$ & 0 & $\cdots$ & $\cdots$ & $\cdots$\\
\hline
IV(a) & $3$ & $2$ & 0 & 0.57186 & 0.00839 & 0.31606\\
IV(a) & $3$ & $2$ & 1 & 0.76898 & 0.01460 & 0.25305\\
IV(a) & $3$ & $2$ & 2 & 0.98177 & 0.02302 & 0.20368\\
IV(b) & $3$ & $2$ & 0 & 0.87934 & 0.01566 & 0.33867\\
IV(c) & $3$ & $2$ & 0 & $\cdots$ & $\cdots$ & $\cdots$\\
IV(d) & $3$ & $2$ & 0 & $\cdots$ & $\cdots$ & $\cdots$\\
\end{tabular}
\end{ruledtabular}
\label{tab:qnm_spp150}
\end{center}
\end{table}

The boundary condition can be imposed in the same manner as that in
the cylindrical model.  We explain the derivation of a boundary
condition at the center in Appendix.  In summary, the regularity
condition at the center can be written as
\begin{equation*}
\delta U_{lm} = C_{lm}^1 r^{\max (\Re[\lambda^{(1)}_{lm}], \Re[\lambda^{(2)}_{lm}])},
\qquad
\delta \Phi_{lm} = C_{lm}^2 r^l,
\end{equation*}
where $\lambda^{(1)}_{lm}$ and $\lambda^{(2)}_{lm}$ are the solutions
of Eq.~(\ref{eqn:lambda}), and $C_{lm}^1$ and $C_{lm}^2$ are
constants.  A surface boundary condition is given as $\delta \Phi_{lm}
= C_{lm}^3 r^{-(l+1)}$, where $C_{lm}^3$ is a function of $C_{lm}^1$
and $C_{lm}^2$ (enable to apply the same technique to meet the surface
boundary condition as in the cylindrical model), and the condition for
$\delta U_{lm}$ is
\begin{equation}
\delta h_{lm} + \xi_{lm}^{j} \nabla_{j} h = 0,
\label{eqn:bc_sp}
\end{equation}
on the surface.  We write down a concrete boundary condition along the
equatorial plane in spherical coordinates as
\begin{eqnarray*}
&&\delta U_{lm} - \delta \Phi_{lm} - \frac{1}{D} (\nabla_{r} \delta U_{lm} ) 
(\nabla_{r} U - \nabla_{r} \Phi) \\
&&
\hspace{1cm}
+ \frac{2m\Omega}{\varpi \tilde{\omega} D} 
(\nabla_{r} U - \nabla_{r} \Phi) \delta U_{lm} = 0,
\end{eqnarray*}
imposing a planner symmetry across the equatorial plane for an
equilibrium configuration.  In conclusion, the system is also set as
an eigenvalue problem.

\subsection{Reflection waves
\label{subsec:reflection}}
In Secs.~\ref{subsec:cylinder} and \ref{subsec:spherical}, we
formulate the stability analysis by finding complex eigenmodes.  Here
we introduce another approach to study the stability of the system: a
wave amplification by inserting incoming waves from the surface to
corotation.  Although the basic idea has already been given in
Ref.~\citep{YS17}, here we write down our techniques, which are useful
for computing large parameter sets and comparing the results with
those of numerical simulations.  We rewrite the basic equation of the
scalar potential $\delta U_m$ [Eq.~(\ref{eqn:Basic_dU_cy})] to focus
on the nature of a wave propagation as \citep{TL08}
\begin{eqnarray}
&&
\left[
\frac{d^2}{d\varpi^2} - V_m^{\rm eff}(\varpi)
\right]
\delta \eta_m(\varpi)
= - \frac{D}{dp/d\rho} S^{-1/2} \delta\Phi_m(\varpi)
,\nonumber \\
&&
\label{eqn:reflection}
\end{eqnarray}
where
\begin{eqnarray*}
S &\equiv& \frac{D}{\rho \varpi}, 
\hspace{1cm}
\delta \eta_m \equiv S^{-1/2} \delta U_m, \\
V_m^{\rm eff}(\varpi) &\equiv &
\frac{D}{dp/d\rho} + \frac{m^2}{\varpi^2} + 
\frac{2m\Omega}{\varpi \tilde{\omega}}
\left( \frac{d}{d\varpi} \ln \frac{\rho \Omega}{D} \right) 
\\
&&
\hspace{1cm}
- S^{1/2}\frac{d^2}{d^2\varpi} S^{-1/2}
.
\end{eqnarray*}
Note that the quantity $V_m^{\rm eff}$ is regarded as an effective
potential of the wave propagation and contains $\tilde{\omega}^{-1}$
term.  Although it is useful to introduce an effective potential to
understand the mechanism of a corotation amplification, the solution
$\delta\eta_m$ contains an apparent singular behavior at Lindbald
radius (the radius where $D=0$).  Therefore, we first construct the
solution of Eq.~(\ref{eqn:reflection}) using a scalar potential
$\delta U_m$, which does not contain a singular behavior at Lindbald
radius, and then transfer the scalar potential $\delta U_m$ to the
perturbed quantity $\delta \eta_m$ to avoid the apparent singular
behavior.

\begin{figure}
\centering
\includegraphics[keepaspectratio=true,width=8cm]{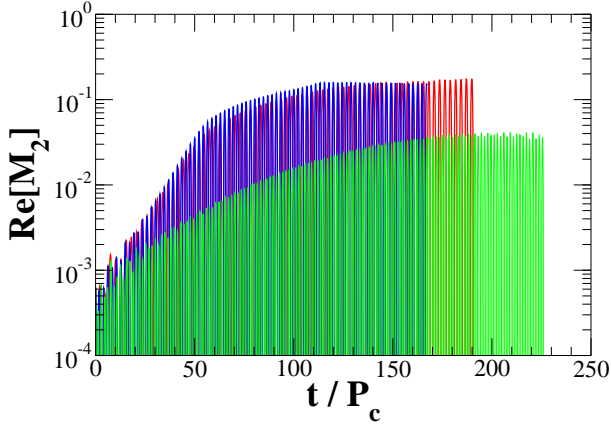}
\caption{
The $M_2$ diagnostics for three low $T/W$ $n=1$ dynamically unstable
stars.  Red, blue, and green lines, respectively, represent models
I-i(a), I-ii(a), and I-iii(a).  We find an amplified oscillation in
the diagnostics for all three cases.
\label{fig:vdig}
}
\end{figure}

\begin{figure}
\centering
\includegraphics[keepaspectratio=true,width=8cm]{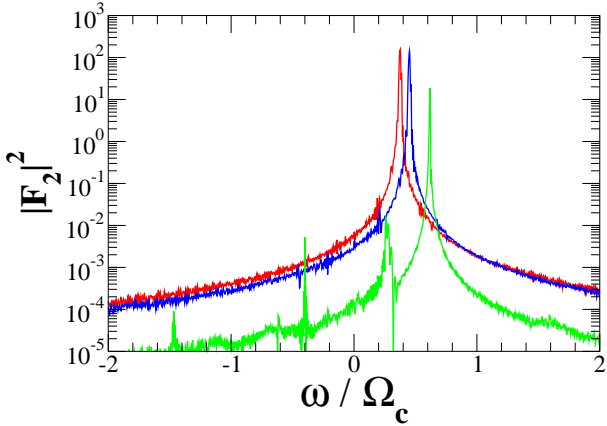}
\caption{
Spectra of $M_2$ diagnostics ($|F_2|^2$) for three low $T/W$ $n=1$
dynamically unstable stars.  Red, blue, and green lines, respectively,
represent models I-i(a), I-ii(a), and I-iii(a).  Only a single
dominant peak in the spectrum shows that the unstable stars contain
one dominant characteristic frequency.
\label{fig:vdigsp}
}
\end{figure}

The basic equation around corotation ($\tilde{\omega} \approx 0$) can
be expressed by extracting the dominant contribution to the singular
behavior in the effective potential $V_m^{\rm eff}$ as
\begin{equation}
\left[
\frac{d^2}{d^2\varpi} - \frac{2m\Omega}{\varpi \tilde{\omega}}
\left( \frac{d}{d\varpi} \ln \frac{\rho\Omega}{D} \right)
\right] \delta U_m (\varpi) = 0.
\label{eqn:eta_cr1}
\end{equation}
The quantity $\tilde{\omega}$ can be expanded up to the first order of
$\varpi - \varpi_{\rm cr}$ as
\begin{eqnarray*}
\tilde{\omega} &=& \Re[\omega] + i \Im[\omega] - m \Omega\\
&=& - m \left. \frac{d\Omega}{d\varpi}\right|_{\varpi=\varpi_{\rm cr}}
(\varpi - \varpi_{\rm cr}) + i \Im[\omega]\\
&=& \frac{g_{\rm cr} \Re[\omega]}{\varpi_{\rm cr}} (\varpi - \varPi_{\rm cr}),
\end{eqnarray*}
where $\varpi_{\rm cr}$ is the cylindrical radius at corotation,
$g_{\rm cr} \equiv - (\varpi_{\rm cr} / \Omega_{\rm cr})
(d\Omega/d\varpi)_{\varpi =\varpi_{\rm cr}}$, and $\varPi_{\rm cr}
\equiv \varpi_{\rm cr} - i (\varpi_{\rm cr} \Im[\omega])/(g_{\rm cr}
\Re[\omega])$.  Then, Eq.~(\ref{eqn:eta_cr1}) can be written as 
\begin{equation}
  \left[ \frac{d^2}{d^2w} + \frac{\beta_{\rm cr}}{w} \right]
  \delta U_m(w) = 0,
\label{eqn:eta_cr2}
\end{equation}
where 
\begin{eqnarray}
w &=& \varpi - \varpi_{\rm cr} + i \frac{\varpi_{\rm cr}
  \Im[\omega]}{g \Re[\omega]},\\
\beta_{\rm cr} &\equiv& \frac{2}{g_{\rm cr}} \frac{d}{d\varpi} \left. \ln
\frac{\kappa^2}{\rho \Omega} \right|_{\varpi=\varpi_{\rm cr}},
\end{eqnarray}
introducing a complex coordinate $w$ in a replacement of a radial one
$\varpi$.  To construct a solution $\delta U_m$ around corotation
analytically, we perform a coordinate transformation as $w=s^2$.  The
basic equation can then be written as 
\begin{equation}
  \left[
    \frac{d^2}{ds^2} + \frac{1}{s} \frac{d}{ds} +
    \left(4 \beta_{\rm cr} - \frac{1}{s^2} \right)
  \right] \delta \Psi_m(s) = 0,
\end{equation}
where $\delta U_m = s \delta \Psi_m$.  Another coordinate
transformation $s = q / (2i\sqrt{|\beta_{\rm cr}|})$, where
$\beta_{\rm cr}<0$ in our equilibrium case, leads to the basic
equation of $\nu=1$ Bessel functions as 
\begin{equation}
  \left[
    \frac{d^2}{dq^2} + \frac{1}{q} \frac{d}{dq} +
    \left(1 - \frac{1}{q^2} \right)
  \right] \delta \Psi_m(q) = 0, 
\end{equation}
containing two independent solutions $N_1(q)$ and $J_1(q)$.  The
general solutions of $\delta U_m$, using Taylor's expansion around
corotation, can then be written as
\begin{equation}
  \delta U_m (w) =
    A_m^1 w + A_m^2 [-4 + w (\ln |w| + 2 \gamma - 2 \ln 2 - 1)],
\label{eqn:dU_corotation}
\end{equation}
where $\gamma$ is Euler's constant, and $A_m^1$ and $A_m^2$ are
constants.  The term $w\ln |w|$ in Eq.~(\ref{eqn:dU_corotation}) is
the origin of a singular behavior of $\delta U_m$ on corotation in a
cylindrical model.

Next we explain our bridging techniques around corotation.  Since we
are focusing on the unstable solution, we introduce a semi-circular
path in the positive imaginary plane to avoid the corotation
singularity in the real axis.  In fact, only an argument difference
appears when bridging the solution at corotation.  Inserting
$w=\varepsilon_{\rm cr} e^{i(\varphi-\pi)}$ ($\varepsilon_{\rm cr}$ is
the radius and $\varphi$ the angle) in the general solution, a
relation between the solution  inside $\delta U_{m}^-$ and outside
corotation $\delta U_{m}^+$ becomes
\begin{widetext}
\begin{align}
\delta U_{m}^- (\varpi) = &
A_m^1 \varpi + A_m^2
[-4 + \varpi (\ln |\varpi| + 2 \gamma - 2 \ln 2 - 1)]
& {\rm as} \quad \varpi < \varpi_{\rm cr}, 
\label{eqn:bridge_dU-}\\
\delta U_{m}^+ (\varpi) =&
A_m^1 \varpi + A_m^2
[-4 + \varpi (\ln |\varpi| - i \pi + 2 \gamma - 2 \ln 2 - 1)]
& {\rm as} \quad \varpi > \varpi_{\rm cr}
\label{eqn:bridge_dU+}.
\end{align}
\end{widetext}

\begin{figure}
\centering
\includegraphics[keepaspectratio=true,width=8cm]{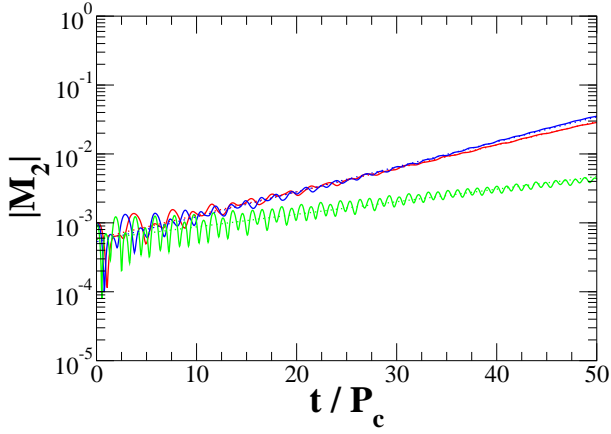}
\caption{
Growth rate of $|M_2|$ diagnostics for three low $T/W$ $n=1$
dynamically unstable stars.  Red, blue, and green lines, respectively,
represent models I-i(a), I-ii(a), and I-iii(a).  Dotted lines,
respectively, represent the fitting formula $|M_2| = A_2
\exp[B_2~t/P_c]$ for each model ($[A_2,B_2] = [5.90 \times 10^{-3},
  7.82 \times 10^{-2}]$ for model I-i[a], $[5.31 \times 10^{-3}, 8.35
  \times 10^{-2}]$ for model I-ii[a], and $[5.74 \times 10^{-3}, 4.20
  \times 10^{-2}]$ for model I-iii[a]).
\label{fig:vdigtime}
}
\end{figure}

\begin{figure}
\centering
\includegraphics[keepaspectratio=true,width=8cm]{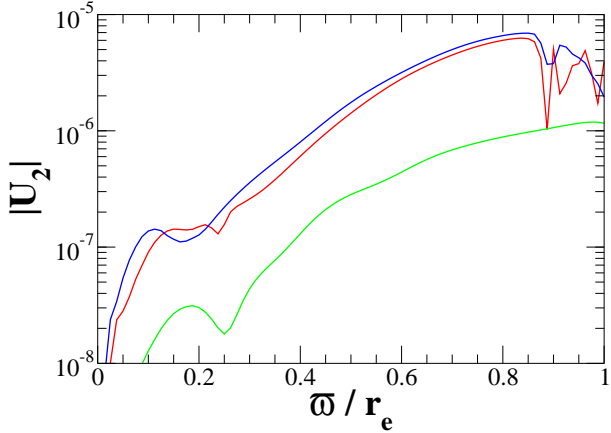}
\caption{
Scalar potentials $|U_2|$ for three low $T/W$ $n=1$ dynamically
unstable stars in the equatorial plane.  Red, blue, and green lines,
respectively, represent models I-i(a) at $t=105.6 P_c$, I-ii(a) at
$t=88.5 P_c$, and I-iii(a) at $t=102.1 P_c$.
\label{fig:vspoten}
}
\end{figure}

\begin{table}[!tb]
\begin{center}
\caption{
The $m=2$ unstable $f$ mode of $n=1$ differentially rotating stars for
three different approaches.}
\begin{ruledtabular}
\begin{tabular}{c c l l l}
Model & Approach & $\Re [\omega] / \Omega_c $ & $\Im [\omega] /
\Omega_c $ & $r_{\rm cr} / r_e$
\\
\hline
I(i-a) & Cylinder & 0.29017 & 0.00855 & 0.48549\\
I(i-a) & Spheroid & 0.35170 & 0.01684 & 0.43297\\
I(i-a) & Simulation & 0.380 & 0.0124 & 0.412\\
\hline
I(ii-a) & Cylinder & 0.36611 & 0.00975 & 0.42251\\
I(ii-a) & Spheroid & 0.43019 & 0.01473 & 0.38205\\
I(ii-a) & Simulation & 0.458 & 0.0133 & 0.367\\
\hline
I(iii-a) & Cylinder & 0.51473 & 0.00862 & 0.33974\\
I(iii-a) & Spheroid & 0.60222 & 0.00936 & 0.30470\\
I(iii-a) & Simulation & 0.622 & 0.00668 & 0.296\\
\end{tabular}
\end{ruledtabular}
\label{tab:validity}
\end{center}
\end{table}

\begin{table}[!tb]
\begin{center}
\caption{
Equilibrium configuration of $n=1$ differentially rotating stars for
verification.}
\begin{ruledtabular}
\begin{tabular}{c r c}
Model & $\Omega_c / \Omega_e$ & $T/W$\\
 \hline
I-i(a) & 26.0 & $6.09 \times 10^{-2}$\\
I-ii(a) & 26.0 & $3.95 \times 10^{-2}$\\
I-iii(a) & 26.0 & $1.90 \times 10^{-2}$\\
\end{tabular}
\end{ruledtabular}
\label{tab:verify_eq}
\end{center}
\end{table}

We can construct a solution $\delta U_m$ which contains an incoming
and a reflection wave outside corotation as follows.  First we impose
a regularity condition at the center for $\delta U_m$ and $\delta
\Phi_m$ [Eq.~(\ref{eqn:cyl_bdcentre})], and we solve a pair of
second-order ordinary differential equations
[Eqs.~(\ref{eqn:Basic_dU_cy}) and (\ref{eqn:Basic_dPhi_cy})] up to
corotation.  Using an analytical bridging technique only for $\delta
U_m$ (computing the two constants $A_m^1$ and $A_m^2$ inside
corotation from $\delta U_m^-$ and $d \delta U_m^- / d\varpi$ and
construct $\delta U_m^+$ and $d \delta U_m^+ / d\varpi$ from these
constants), we are able to solve continuously a pair of second-order
ordinary differential equations [Eqs.~(\ref{eqn:Basic_dU_cy}) and
  (\ref{eqn:Basic_dPhi_cy})] from corotation up to the surface.  Note
that we adopt the same technique to impose a boundary condition for
$\delta\Phi_m$ on the surface ($\delta \Phi_m = C_m^3 \varpi^{-|m|}$)
as in Sec.~\ref{subsec:cylinder}.

After constructing a solution of the perturbed quantity $\delta
\eta_m$ from a scalar potential $\delta U_m$, here we explain our
method to extract the reflection amplitude.  The solution of a
perturbed quantity $\delta \eta_m$ in the wave propagation region can
be explained as
\begin{equation}
\delta \eta_m (\varpi) = I_m (\varpi) e^{-ik_m \varpi} + R_m(\varpi)
e^{ik_m \varpi},
\end{equation}
where $k_m \equiv \sqrt{-V_m^{\rm eff}}$.  Using $\delta \eta_m$ and
$d\delta\eta_m / d\varpi$ around the surface, an amplitude of incoming
and outgoing waves to corotation  around the surface can be extracted
as
\begin{eqnarray}
  I_m(\varpi) &=&
  \frac{1}{2} e^{ik_m\varpi}
  \left(
    \delta \eta_m - \frac{1}{ik_m} \frac{d \delta \eta_m}{d\varpi}
  \right), \\
  R_m(\varpi) &=&
  \frac{1}{2} e^{-ik_m\varpi}
  \left(
    \delta \eta_m + \frac{1}{ik_m} \frac{d \delta \eta_m}{d\varpi}
  \right).
\end{eqnarray}

\begin{figure*}
\centering
\includegraphics[keepaspectratio=true,width=16cm]{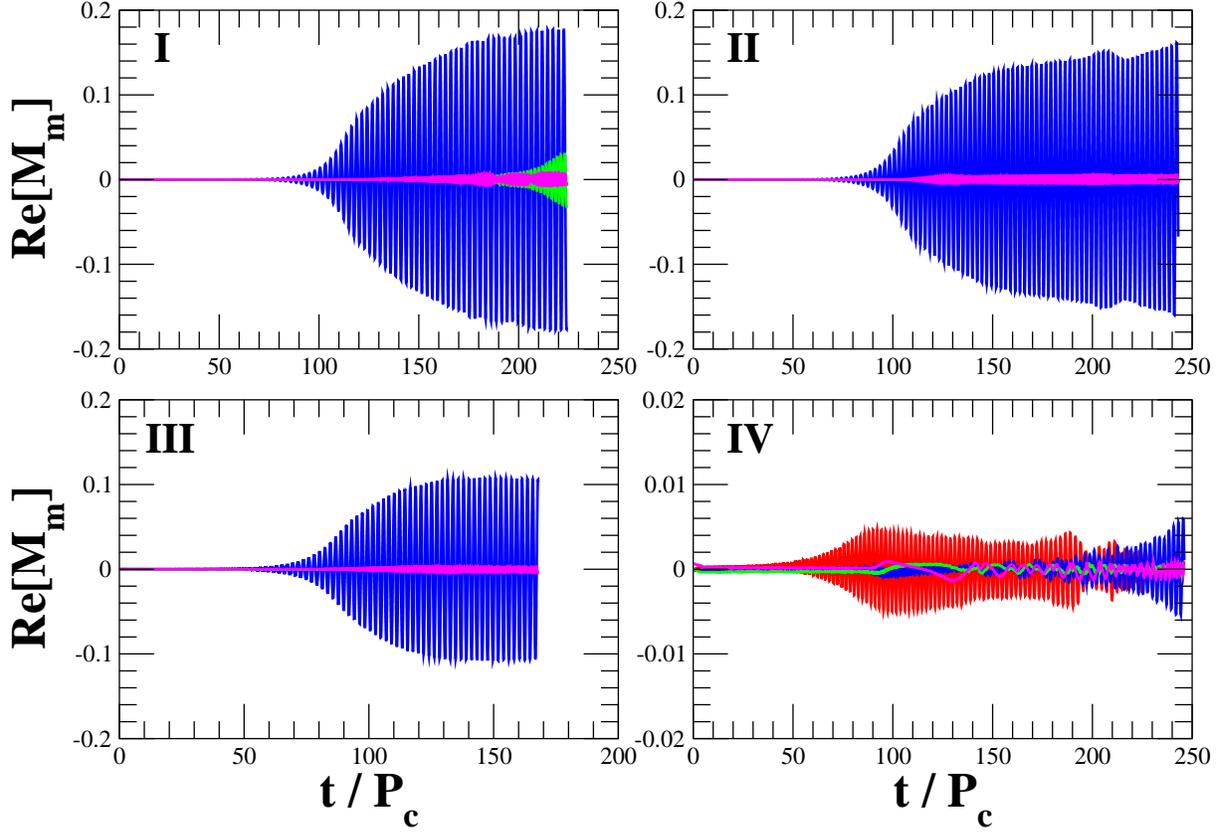}
\caption{
The $M_m$ diagnostics for four low $T/W$ dynamically unstable stars.
Red, blue, green, and magenta lines, respectively, represent
diagnostics $M_1$, $M_2$, $M_3$, and $M_4$.
\label{fig:dig}
}
\end{figure*}

The growth timescale can be interpreted as an amplification of sound
waves by the corotation barrier through a single reflection cycle of
the waves.  Suppose a perturbed quantity $\delta \eta_m$ grows
exponentially in time in the wave propagation regime as
\begin{widetext}
\begin{eqnarray}
\delta \eta_m(t,\varpi) &=& e^{-i \omega_{\rm ref} t} 
\left(
  I_m (\varpi) e^{-ik_m \varpi} + R_m(\varpi) e^{ik_m \varpi}
\right)
\nonumber \\
&=&
e^{\Im[\omega_{\rm ref}] t} 
\left(
  I_m (\varpi) e^{-i(\Re[\omega_{\rm ref}] t + k_m \varpi)} + 
  R_m(\varpi) e^{-i(\Re[\omega_{\rm ref}] t - k_m \varpi)}
\right),
\end{eqnarray}
\end{widetext}
where $\omega_{\rm ref}$ is a complex frequency including a growth
timescale in the imaginary part, illustrating a wave amplification due
to the corotation barrier.  The amplification rate through a single
wave reflection by the effective potential is $|R_m| / |I_m|$.  Once
we introduce a wave-traveling time $T_m$ through one reflection by the
potential, a relation between the amplification factor and the
imaginary part of the frequency is
\begin{equation}
\frac{|R_m|}{|I_m|} = \exp\left[ \Im[\omega_{\rm ref}] T_m \right],
\label{eqn:amp}
\end{equation}
where the wave-traveling time $T_m$ is computed as
\begin{equation}
T_m = \frac{2}{\Re[\omega_{\rm ref}]}\int_{\varpi_{\rm
    Vmin}}^{\varpi_{\rm Vmax}} k_m d\varpi.
\end{equation}
Note that the radii $\varpi_{\rm Vmin}$ and $\varpi_{\rm Vmax}$(or the
equatorial surface radius $r_e$ if there is no such radius) represent
the turning point (the radius where $V_m^{\rm eff} = 0$) outside
corotation, and the frequency $\Re[\omega_{\rm ref}]$ is taken from
the real part of the eigenfrequency by the normal mode analysis
computed in Sec.~\ref{subsec:cylinder}.  The imaginary part of the
frequency $\Im[\omega_{\rm ref}]$ is derived from Eq.~(\ref{eqn:amp})
as
\begin{equation}
  \Im[\omega_{\rm ref}] =
  \frac{\ln |R_m| - \ln |I_m|}{T_m} =
  \frac{\Re[\omega_{\rm ref}] (\ln |R_m| - \ln |I_m|)}
       {2 \int_{\varpi_{\rm Vmin}}^{\varpi_{\rm Vmax}} k_m d\varpi}
,
\label{eqn:timescale_ref}
\end{equation}
extracting the amplitude of inserted and amplified waves at the radius
$\varpi = \varpi_{\rm Vmax}$.

\begin{figure*}
\centering
\includegraphics[keepaspectratio=true,width=16cm]{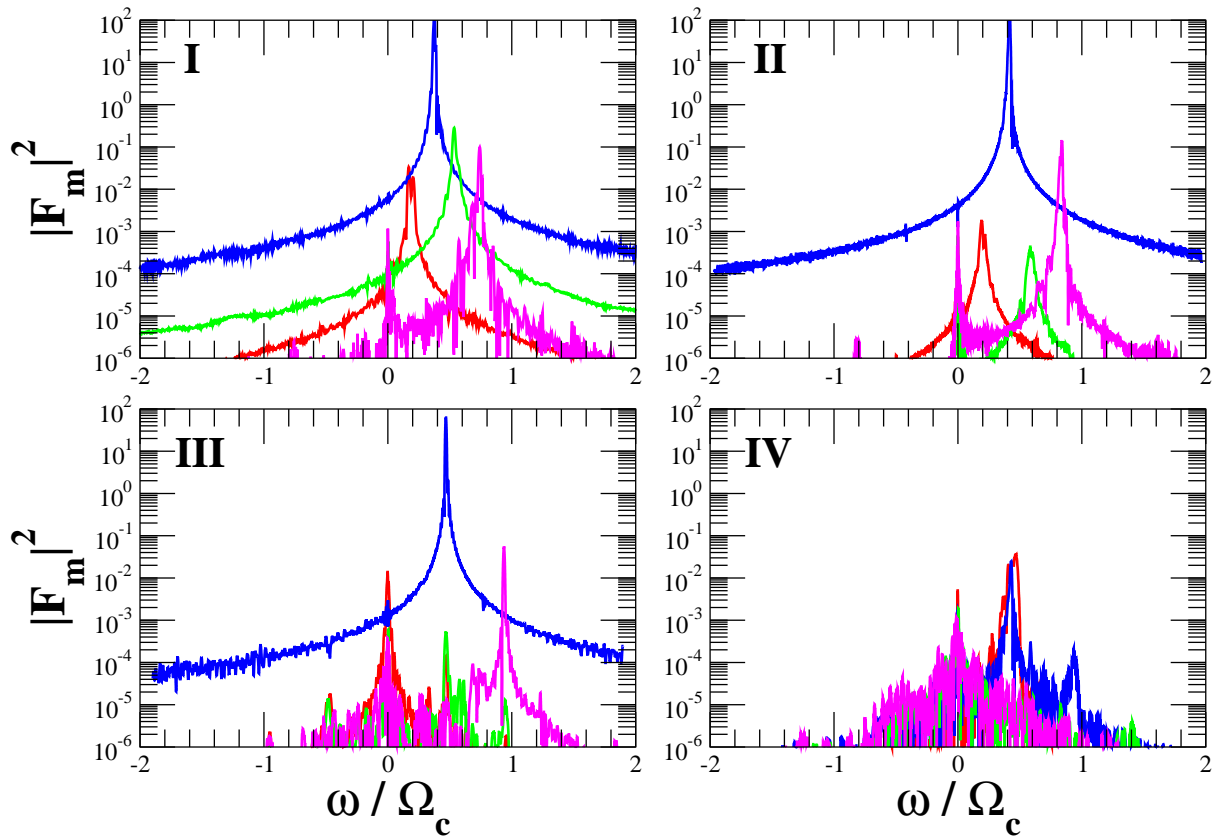}
\caption{
Spectra of $M_m$ diagnostics ($|F_m|^2$) for four low $T/W$
dynamically unstable stars.  Red, blue, green, and magenta lines
respectively represent $m=1$, $2$, $3$, and $4$.  Only a single
dominant peak in the spectrum shows that the unstable stars contain
one dominant characteristic frequency.
\label{fig:digsp}
}
\end{figure*}

\subsection{Stability analysis
\label{subsec:stability}}
The axisymmetric equilibrium configuration of the differentially
rotating stars is computed in the two-dimensional cylindrical
coordinates \citep{SK08}.  Then we take the equilibrium quantities,
the pressure over rest mass density $q$ ($\equiv p/\rho$) and
gravitational potential $\Phi$, 3841 grid points uniformly along the
equatorial plane from the center to the stellar surface in order to
integrate the pulsation equations.  We use the fourth-order
Runge-Kutta method (fourth-order integration in space) to integrate
Eqs.~(\ref{eqn:Basic_dU_cy}) and (\ref{eqn:Basic_dPhi_cy}) for
cylindrical models and Eqs.~(\ref{eqn:Basic_dU_sp}) and
(\ref{eqn:Basic_dPhi_sp}) for spheroidal models with both 1921 grid
points in the normal mode analysis.

We search the complex frequency $\omega$ in the region of $\Re[\omega]
\in [0,4]\Omega_c$ and $\Im[\omega] \in [0,0.2]\Omega_c$, integrating
Eqs.~(\ref{eqn:Basic_dU_cy}) and (\ref{eqn:Basic_dPhi_cy}) for
cylindrical models or Eqs.~(\ref{eqn:Basic_dU_sp}) and
(\ref{eqn:Basic_dPhi_sp}) for spheroidal models from the center to the
surface to check whether the boundary condition of
Eq.~(\ref{eqn:bc_cy}) for cylindrical models or Eq.~(\ref{eqn:bc_sp})
for spheroidal models is satisfied.  Note that the frequency we search
covers the region where corotation exists inside the stars.  In fact,
we compute the left-hand side of Eq.~(\ref{eqn:bc_cy}) for cylindrical
models or Eq.~(\ref{eqn:bc_sp}) for spheroidal models normalized by
$\delta h_m$ or $\delta h_{lm}$ for each complex frequency, and we
determine the eigenfrequency once the following two conditions are
satisfied.  The first is that the relative error of the left-hand side
of Eq.~(\ref{eqn:bc_cy}) for cylindrical models or
Eq.~(\ref{eqn:bc_sp}) for spheroidal models normalized by $\delta h_m$
or $\delta h_{lm}$ is less than $2\times 10^{-3}$.  The second is that
the frequency takes the minimum around the neighboring four complex
frequencies of each grid point in the complex plane.  Our frequency
resolution for finding the eigenfrequency is $\Delta \omega = 1 \times
10^{-5}$ for both real and imaginary parts.  We only focus on $m=1$
spiral and $m=2$ bar mode here.

We show the eigenfrequencies from the linear perturbation analysis in
Tables~\ref{tab:qnm_cyp150-1} -- \ref{tab:qnm_cyp150-4} for
cylindrical models and in Table~\ref{tab:qnm_spp150} for spheroidal
models.  We also show the eigenfunctions in Fig.~\ref{fig:efd02p150cy}
for cylindrical models and in Fig.~\ref{fig:efd02p150sp} for
spheroidal models.  As we have already stated in Ref.~\citep{SY16}, we
find the following three conclusions.  One is that an oscillation
between corotation and the surface can be found in all eigenfunctions.
Every node number can be seen in all differentially rotating stars for
$m=2$ bar mode.  Note that some of the zero nodes for $m=1$, which
represents the shift of the center of mass, correspond to the pure
imaginary eigenfrequency (we omit these pure imaginary
eigenfrequencies from the tables).  Another is that all
eigenfrequencies that have a corotation inside the star are unstable
(existence of a positive imaginary part in eigenfrequencies).  This
fact indicates that the existence of corotation triggers dynamical
instabilities.  The other is that $m=1$ dynamical instabilities become
dominant in a soft equation of state.  This means that we are able to
identify the stiffness of the equation of state by the direct
detection of gravitational waves.

\begin{figure}
\centering
\includegraphics[keepaspectratio=true,width=8cm]{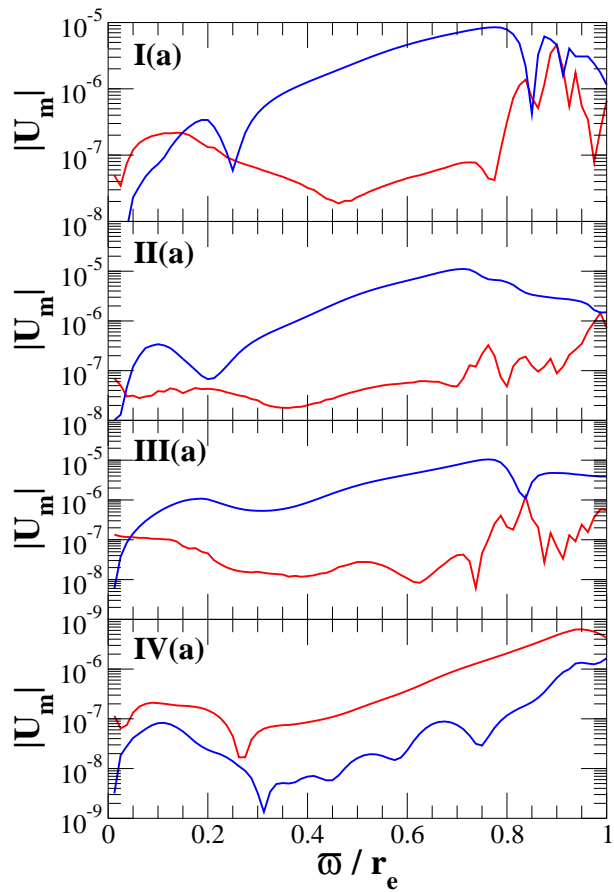}
\caption{
Scalar potentials of $|U_1|$ (red line) and $|U_2|$ (blue line)
diagnostics for four low $T/W$ dynamically unstable stars in the
equatorial plane.  The potentials are respectively plotted at
$t=171.8 P_c$ (model I[a]), at $t=219.3 P_c$ (model II[a]), at
$t=106.75 P_c$ (model III[a]), and at $t=78.11 P_c$ (model IV[a]).
\label{fig:ef}
}
\end{figure}

For the reflection wave analysis, we basically adopt the same
computational technique as in the normal mode analysis.  We take again
the equilibrium quantities, the pressure over rest mass density $q$
and gravitational potential $\Phi$, 3841 grid points uniformly along
the equatorial plane, and use the fourth-order Runge-Kutta method
(fourth-order integration in space) to integrate
Eqs.~(\ref{eqn:Basic_dU_cy}) and (\ref{eqn:Basic_dPhi_cy}) for
cylindrical models with both 1921 grid points in the normal mode
analysis.  Only the difference of computational techniques from the
normal mode analysis is an introduction of bridging of the perturbed
scalar potential $\delta U_m$ around corotation due to a coordinate
singularity at corotation.  We assume that the corotation is not
located on the grid point (the measure is zero in a mathematical
sense), and separate two regions as inside and outside corotation.
Bridging the perturbed scalar function $\delta U_m$ between two
different regions,  two constants $A_m^1$ and $A_m^2$ at the closest
inner grid point to corotation are used to construct the solution.  We
summarize our finding of the reflection timescale (corresponds to an
imaginary part of the reflection frequency), which is based on the
idea of an amplification mechanism through a corotation singularity,
in Tables~\ref{tab:qnm_cyp150-1} -- \ref{tab:qnm_cyp150-4} for
cylindrical models.  Although estimating the timescale of a single
reflection wave contains many crude approximations (assumption of a
wave propagation in the equatorial plane, a transmitting wave is not
taken into account for deriving timescale $T_m$ and estimating
wave-traveling time is assumed by the time of one reflection from the
potential), qualitative features seem to be similar to those of the
normal mode analysis. Although a complete understanding of the
mechanism requires a two-dimensional nonaxisymmetric perturbation
analysis, which is out of our scope in this paper, our finding may
enlighten a direction to understand a physical mechanism for low $T/W$
dynamical instabilities.

\subsection{Picture of low $T/W$ dynamical instabilities
\label{subsec:picture}}
In Sec.~\ref{subsec:reflection}, we qualitatively have good agreement
between the stability analysis of a scattering problem and a normal
mode analysis.  We are able to propose the following mechanism for
generating low $T/W$ dynamical instabilities.  Suppose that an
eigenfrequency of a pulsation mode, such as $f$ or $p$ mode, shows an
existence of corotation inside the star.  The mode grows exponentially
in a nonaxisymmetric manner due to an amplification mechanism.  After
the angular momentum transport efficiently plays a role due to
nonaxisymmetric deformation, the amplification condition may no longer
be satisfied.  As a result, the growth of the instabilities are at
least saturated.  If this picture is correct, all existing pulsation
modes are the potential candidates to unstabilize the system when the
amplification condition sets in.  However, the eigenfrequencies that
contain corotation inside the star are quite limited in rotating
equilibrium stars.  Also, a certain degree of differential rotation is
required.  In addition, the growth timescale depends on the
configuration of an effective potential, which is normally powerful to
the $f$ mode.  Finally, a saturation amplitude depends on the
efficiency of the angular momentum transport in the instabilities.

\section{Numerical results
\label{sec:results}}
\subsection{Validity of cylindrical and spheroidal models
\label{subsec:validity}}

\begin{figure*}
\centering
\includegraphics[keepaspectratio=true,width=16cm]{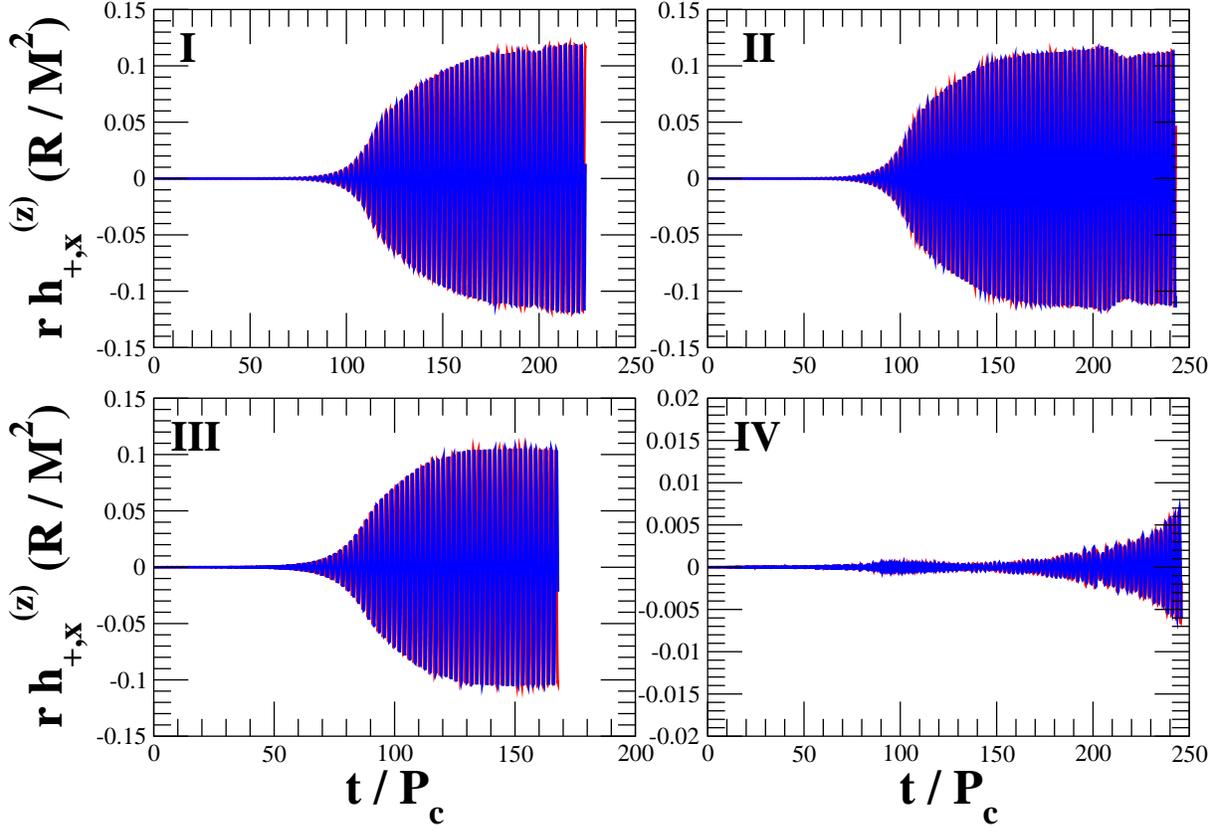}
\caption{
Gravitational waveforms for four low $T/W$ dynamically unstable stars
observed along the rotational axis of the equilibrium star.  Red and
blue lines, respectively, represent $+$ and $\times$ modes.
\label{fig:gwz}
}
\end{figure*}

\begin{figure*}
\centering
\includegraphics[keepaspectratio=true,width=16cm]{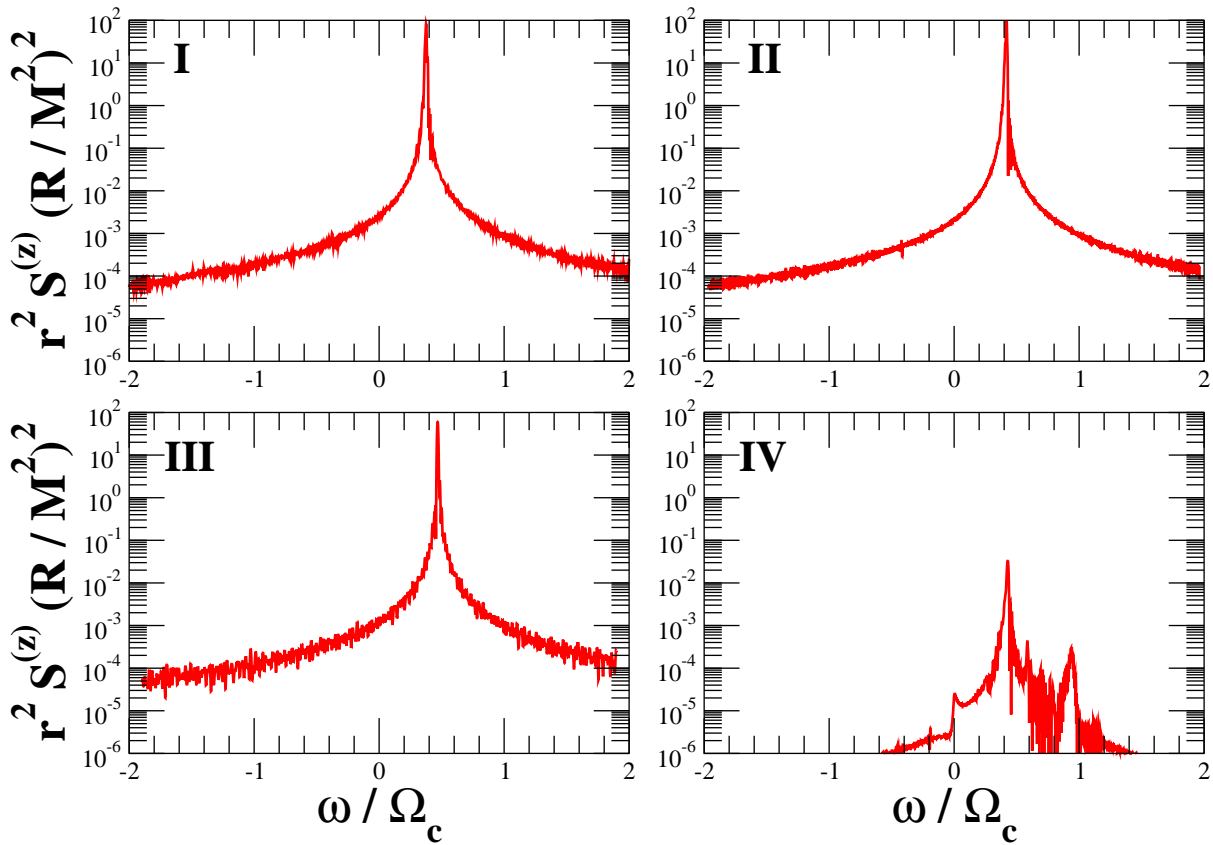}
\caption{
Spectra of gravitational waveforms observed along the rotational axis
for four low $T/W$ dynamically unstable stars.
\label{fig:gwspkz}
}
\end{figure*}

We briefly introduce our results of three-dimensional hydrodynamical
simulations in Newtonian gravity and compare them with those of linear
perturbative analyses.  Here we choose three differentially rotating
equilibrium stars, keeping the same polytropic index $n=1$ and degree
of differential rotation $\Omega_e / \Omega_c = 26.0$, where
$\Omega_c$ is the central angular velocity of the star and $\Omega_e$
the equatorial surface angular velocity, but varying the deformation
rate $1-r_p / r_e$ as $0.125, 0.250, 0.375$, where $r_p$ is the polar
surface radius of the star, summarized in Table~\ref{tab:verify_eq}.
We impose a nonaxisymmetric perturbation in the rest mass density as 
\begin{equation*}
\rho = \rho_{\rm eq}\left( 1 + \delta \frac{x^2 + 2 x y - y^2}{r_e^2}
\right),
\end{equation*}
where we set $\delta = 5 \times 10^{-3}$ for evolution.  Note that
$\rho_{\rm eq}$ is an equilibrium configuration of the rest mass
density, $x$ and $y$ are the components of Cartesian coordinates.
Note that 161 grid points are covered along the equatorial diameter of
the star, with twice equatorial radius as large as the outer boundary
for each coordinate direction.  We monitor the diagnostics $M_2$ as 
\begin{equation}
M_m = \langle e^{im\varphi}\rangle_{m} = \frac{1}{M} \int dv~\rho
e^{im\varphi},
\end{equation}
which is $m=2$ rest mass density weighted average in the whole volume,
and find that $M_2$ grows exponentially for low $T/W$ dynamically
unstable case (Fig.~\ref{fig:vdig}).  In practice, the $M_2$ grows 
exponentially up to $t \approx 50~P_c$ for models I-i(a) and I-ii(a),
and to $t \approx 150~P_c$ for model I-i(c), and saturates its
amplitude around $M_2 \approx 0.1$ for models I-i(a) and II-i(a), and
around $M_2 \approx 0.04$ for model III-i(a).  We extract the
characteristic frequencies in the diagnostics from computing their
spectra in Fig.~\ref{fig:vdigsp} as
\begin{equation}
  |F_m|^2 =
  \left|
    \frac{1}{M} \int dt \int dx^3 \rho e^{i(\omega t - m \varphi)}
  \right|^2.
\end{equation}
The peak frequencies for models I-i(a), I-ii(a), and I-iii(a) are,
respectively, $\omega / \Omega_c = 0.380$, $0.458$, and $0.622$.  Each
model contains a single peak in the positive frequency region,
indicating that our three models contain one dominant unstable
``mode.''  The growth time of the diagnostics
(Fig.~\ref{fig:vdigtime}) is extracted by using a fitting formula of
the exponential growth function as
\begin{equation}
M_m = A_m \exp \left[ B_m t / P_c \right],
\end{equation}
where $A_m$ and $B_m$ are the two dimensionless parameters to be
fitted for each model.  We show a clear fitting to the diagnostics
curve in Fig.~\ref{fig:vdigtime} with the choice of parameter sets
described in the caption.  The imaginary parts of the characteristic
frequencies are extracted by using the relation $\Im[\omega]=B_m/(2
\pi)$  as $\Im[\omega]/\Omega_c=0.0124$ for model I-i(a), $0.0133$ for
model I-ii(a), and $0.00668$ for model I-iii(a). 

We also monitor a scalar potential $U_m$ in the simulation
\citep{SY16}, which represents the eigenfunction of rotating pulsating
stars in Newtonian gravity, as
\begin{equation*}
U_{m} = \frac{1}{2 \pi U} \int d\varphi~ue^{im\varphi}, 
\hspace{5mm}
U = \int_V dv~u, 
\end{equation*}
where $u\equiv H + \Phi=\varepsilon + p / \rho + \Phi$
(Fig.~\ref{fig:vspoten}).  In all three low $T/W$ $n=1$ dynamically
unstable stars, the scalar potential contains a single local minimum
around $\varpi / r_e \approx 0.2$--$0.3$, which may express a singular
behavior at corotation in pulsation equations.  Only a monotonic
increase of the potential between corotation and the surface indicates
that the dominant frequency we find in simulations represents a
fundamental ($f$) mode in rotating pulsating stars.

Our comparison of characteristic complex frequencies (which represent
the oscillation frequencies and the growth times) between three
different approaches --- a cylindrical model, a spheroidal model, and
a numerical simulation --- is summarized in Table~\ref{tab:validity}.
As deformation of the equilibrium star becomes small, the results of a
spheroidal model and a numerical simulation approach each other.  This
feature is especially seen in the corotation radius of the star and
the growth rate of the instabilities.  In a spherically symmetric
background, it is natural to expand the perturbed quantities using a
spherical harmonics.  As a star deviates from a spherical symmetry, a
spheroidal model becomes only an approximation, since the model
assumes spherical harmonic dependence.  In addition, there is also
good agreement of the results between cylindrical and spheroidal
models when the deformation rate is small.  Since a cylindrical model
has less restriction for computing the normal modes in the frequency
regions, we mainly explore the results of a cylindrical model in a
wide frequency regime and present generic features in
Sec.~\ref{subsec:simulations}.

\begin{figure*}
\centering
\includegraphics[keepaspectratio=true,width=16cm]{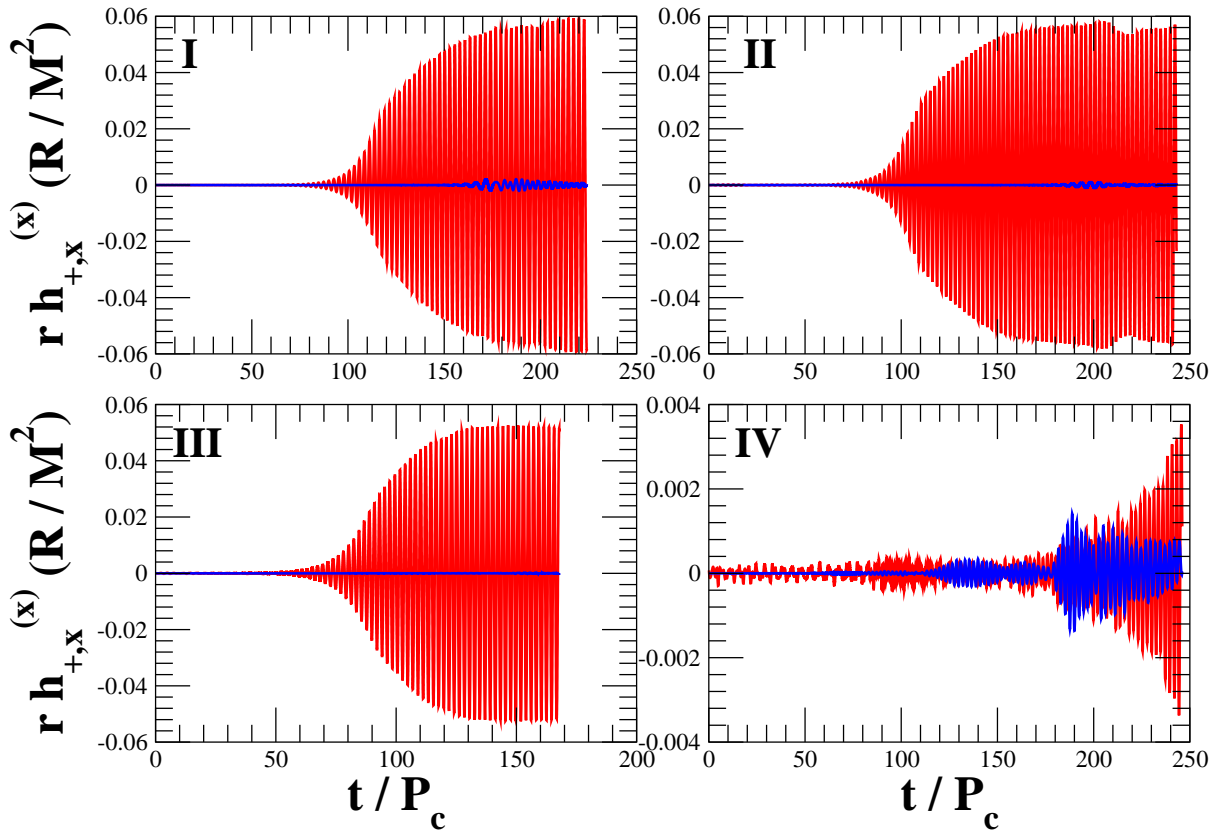}
\caption{
Same as Fig.~\ref{fig:gwz}, but along the principal axis in the
equatorial plane.
\label{fig:gwx}
}
\end{figure*}

\subsection{Numerical simulations
\label{subsec:simulations}}
We pick up four low $T/W$ dynamically unstable stars (I[a], II[a],
III[a], and IV[a] of Table \ref{tab:equilibrium}), varying the
stiffness of the equation of state, to focus on the dominancy of
spiral ($m=1$) and bar ($m=2$) modes throughout the evolution.  Here
we impose the following density perturbation in the equilibrium
configuration as:
\begin{equation*}
\rho = \rho_{\rm eq}
\left( 
  1 + 
  \delta^{(1)} \frac{x + y}{r_e}  + 
  \delta^{(2)} \frac{x^2 + 2 x y - y^2}{r_e^2} 
\right),
\end{equation*}
where we set $\delta^{(1)} =\delta^{(2)} = 1 \times 10^{-5}$ to
trigger the instabilities.  We monitor $M_m$ diagnostics ($M_1$,
$M_2$, $M_3$, and $M_4$) for all four unstable stars, shown in
Fig.~\ref{fig:dig}.  For models I, II, and III, the $m=2$ diagnostics
grow exponentially up to $\approx 0.10$--$0.15$, with substantial
growth of $m=4$.  Especially for model I, the $m=3$ diagnostic grows
exponentially around $t \agt 180 P_c$, which may be explained as a
nonlinear mode coupling from the bar mode investigated in
Refs.~\citep{SK08,KS08}.  In contrast to the former three models,
model IV contains a qualitative difference.  The $m=1$ diagnostic
grows exponentially up to $\approx 0.006$, with substantial growth of
$m=2$--$4$ around $t \agt 150 P_c$.  This feature can be used to
restrict the stiffness of the equation of state.  This subject will be
discussed in Sec.~\ref{subsec:constrainEOS}.

We compute the spectra of the diagnostics $M_m$ in
Fig.~\ref{fig:digsp}.  We find a clear peak for each diagnostic.
Model I has a peak at $\omega = 0.168 \Omega_c$ for $m=1$, $\omega =
0.375 \Omega_c$ for $m=2$, $\omega = 0.535 \Omega_c$ for $m=3$, and
$\omega = 0.742 \Omega_c$ for $m=4$.  Model II has a peak at $\omega =
0.191 \Omega_c$ for $m=1$, $\omega = 0.409 \Omega_c$ for $m=2$,
$\omega = 0.583 \Omega_c$ for $m=3$, and $\omega = 0.825 \Omega_c$ for
$m=4$.  Model III has a peak at  (no peak for $m=1$) $\omega = 0.485
\Omega_c$ for $m=2$, $\omega = 0.469 \Omega_c$ for $m=3$, and $\omega
= 0.937 \Omega_c$ for $m=4$.  From these three models, the $M_2$
diagnostics take the dominant role as the maximum spectrum amplitude
is the highest for all four diagnostics.  Also the peak frequency of
$m=4$ is almost twice that of $m=2$, indicating that they are
generated from the same corotation and act as harmonics.  In contrast,
the odd $M_m$ diagnostics do not have a harmonic behavior to the bar
mode, meaning that they are generated from different corotations.

We show scalar potentials $U_m$ for $m=1$ and $m=2$ in
Fig.~\ref{fig:ef}.  Although the definition of a scalar potential has
been taken from the feature of a perturbative approach, we clearly
find the same behavior even in the results of three-dimensional
simulations, indicating a clear correspondence to the existence of
corotation.

\begin{figure*}
\centering
\includegraphics[keepaspectratio=true,width=16cm]{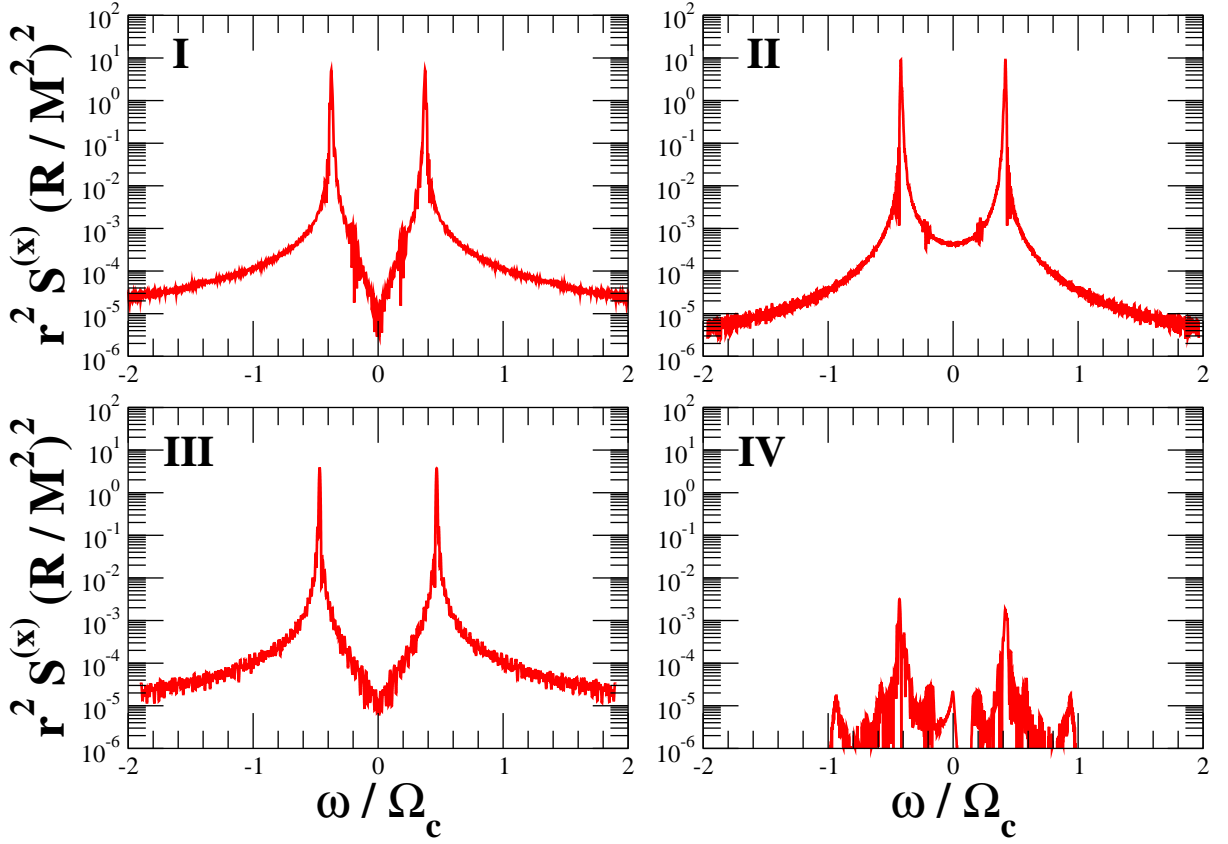}
\caption{
Same as Fig.~\ref{fig:gwspkz}, but in the equatorial plane.
\label{fig:gwspkx}
}
\end{figure*}

\subsection{Saturation amplitude and gravitational waves}
We compute approximate gravitational waveforms by evaluating the
quadrupole formula.  In the radiation zone, gravitational waves can be
described by a transverse-traceless, perturbed metric $h_{ij}^{TT}$
with respect to a flat spacetime. In the quadrupole formula,
$h_{ij}^{TT}$ is found from \citep{MTW}
\begin{equation}
h_{ij}^{TT}= \frac{2}{r} \frac{d^{2}}{d t^{2}} I_{ij}^{TT},
\label{eqn:wave1}
\end{equation}
where $r$ is the distance to the source, where $I_{ij}$ is the
quadrupole moment of the mass distribution [see Eq.~{(36.42b)} in
  Ref.~\citep{MTW}], and where $TT$ denotes the transverse-traceless
projection.  Choosing the direction of the wave propagation to be
along the $x$ axis (one of the principal axes in the equatorial plane
of the equilibrium star) and $z$ axis (rotational axis of the
equilibrium star), we determine the two polarization modes of
gravitational waves from
\begin{eqnarray}
&&h_{+}^{(x)} \equiv \frac{1}{2} (h_{yy}^{TT} - h_{zz}^{TT})
\mbox{~~~and~~~} 
h_{\times}^{(x)} \equiv h_{yz}^{TT}
,\\
&&h_{+}^{(z)} \equiv \frac{1}{2} (h_{xx}^{TT} - h_{yy}^{TT})
\mbox{~~~and~~~} 
h_{\times}^{(z)} \equiv h_{xy}^{TT}.
\end{eqnarray}
For observers along the $x$ axis and $z$ axis, we thus have
\begin{eqnarray}
\frac{r h_{+}^{(x)}}{M} &=& 
\frac{1}{2 M} \frac{d}{d t} (\dot{I}_{yy} - \dot{I}_{zz}),
\label{eqn:h+x}
\\
\frac{r h_{\times}^{(x)}}{M} &=& 
\frac{1}{M} \frac{d}{d t} \dot{I}_{yz}
\label{eqn:h-x}
,\\
\frac{r h_{+}^{(z)}}{M} &=& 
\frac{1}{2 M} \frac{d}{d t} (\dot{I}_{xx} - \dot{I}_{yy}),
\label{eqn:h+z}
\\
\frac{r h_{\times}^{(z)}}{M} &=& 
\frac{1}{M} \frac{d}{d t} \dot{I}_{xy}
\label{eqn:h-z}
.
\end{eqnarray}
Note that $\dot{A}$ represents the time derivative of $A$.  The number
of time derivatives $I_{ij}$ that have to be taken out can be reduced
by using the continuity equation [Eq.~(\ref{eqn:continuity})]
\begin{equation}
\dot{I}_{ij} = \int (\rho v^{i} x^{j} + \rho x^{i} v^{j}) d^{3}x,
\end{equation}
in Eqs.~(\ref{eqn:h+x}), (\ref{eqn:h-x}), (\ref{eqn:h+z}), and
(\ref{eqn:h-z}) (see Ref.~\citep{Finn89}).

The spectrum of a gravitational waveform can be computed as 
\begin{equation}
S =  |\tilde{h}_{+}|^2 + |\tilde{h}_{\times}|^{2},
\end{equation}
where 
\begin{equation}
\tilde{h}_{+, \times} = \int dt h_{+, \times} e^{i \omega t}.
\end{equation}

We show gravitational waveforms (Fig.~\ref{fig:gwz}) and their spectra
(Fig.~\ref{fig:gwspkz}) along the equilibrium rotational axis from
four different low $T/W$ dynamically unstable stars.  We find
quasi-periodic oscillations for all four models for both $+$ and
$\times$ modes.  Also a single characteristic frequency can be seen in
the spectra for all models.  For example, $\omega = 0.375 \Omega_c$
for model I, $\omega = 0.418 \Omega_c$ for model II, $\omega = 0.468
\Omega_c$ for model III, and $\omega = 0.429 \Omega_c$ for model IV.
Since the observer is set along the rotational axis, we can only
observe $m(=l)\gtrsim2$ diagnostics.  Therefore, all frequencies of
peaks in the spectra correspond to those in $m=2$ diagnostics.

In order to focus on the detectability of $m=1$ diagnostics, we next
locate the observer along the equatorial plane of the equilibrium
stars and show gravitational waveforms (Fig.~\ref{fig:gwx}) and their
spectra (Fig.~\ref{fig:gwspkx}).  In this case, all $m$ modes ($-l
\leq m \leq l$ for each $l$ modes) can be observed.  We only find a
quasi-periodic oscillation for all four models for $+$ modes,
indicating the feature of $m=2$ diagnostics.  We also find an
amplified oscillation in $\times$ mode when the $m=1$ diagnostic
grows, indicating one feature of the $m=1$ diagnostic.  This feature
can clearly be seen in the spectrum of gravitational waves.  Models I
and II have two peaks in the spectrum of positive frequency, $\omega =
0.192 \Omega_c$, $0.375 \Omega_c$ for model I and $\omega = 0.218
\Omega_c$, $0.418 \Omega_c$ for model II.  Comparing to the peak
frequencies in the $M_m$ diagnostics, two peak frequencies in the
gravitational waveforms respectively correspond to $m=1$ and $m=2$
diagnostics.  Model III has a single positive frequency $\omega =
0.468 \Omega_c$, which corresponds to the $m=2$ diagnostic.  Model IV
has four peaks in low amplitudes compared to models I, II, and III,
and the peak frequencies are $\omega = 0.190 \Omega_c$, $0.417
\Omega_c$, $0.610 \Omega_c$, and $0.927 \Omega_c$.

\subsection{Constrain the stiffness of the equation of state
\label{subsec:constrainEOS}}
\begin{figure}
\centering
\includegraphics[keepaspectratio=true,width=8cm]{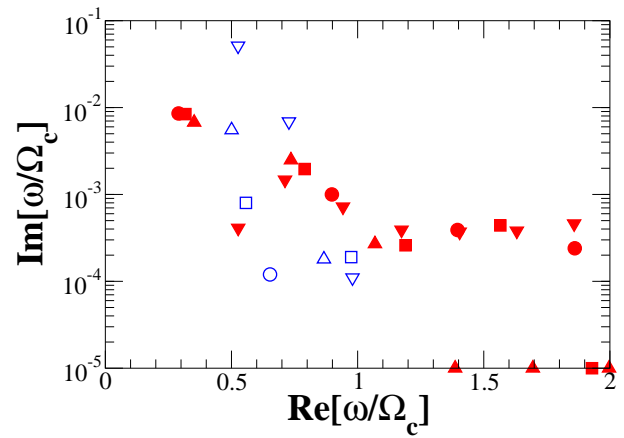}
\caption{
The $m=1$ and $m=2$ eigenfrequencies for four low $T/W$ dynamically
unstable stars (models I[a], II[a], III[a], and IV[a]) in cylindrical
models.  Circles, squares, top triangles, and bottom triangles,
respectively, denote the polytropic index of $n=1$, $1.5$, $2$, and
$3$.  Opened and filled symbols, respectively, represent the $m=1$ and
$m=2$ modes.  Comparing the largest imaginary frequencies between
$m=1$ and $m=2$ in each polytropic index, the dominant $m$ mode
changes at the stiffness of the equation of state around $\Gamma
\approx 1.50$.
\label{fig:dominancy}
}
\end{figure}

Finally we propose one procedure for constraining the stiffness of the
equation of state by the direct detection of gravitational waves.   We
find from three-dimensional numerical simulations that the dominancy
mode between $m=1$ and $m=2$ throughout the evolution strongly depends
on the stiffness of the equation of state (Fig.~\ref{fig:dig}).
Models I, II, and III have $M_2$ diagnostics dominancy, while model IV
has $M_1$ dominancy.  We also extract the growth rate of the
instabilities from each $M_m$ diagnostic using the same fitting
formula as in Sec. \ref{subsec:validity} but changing the starting
time $t_0$ of the exponential growth as
\begin{equation}
M_m = A_m \exp \left[ B_m (t-t_0) / P_c \right].
\end{equation}
Note that we extract two constants $A_m$ and $B_m$ through around 50
central rotation periods of the equilibrium stars from the time
$t_0$.  The imaginary part of complex characteristic frequencies of
model I are $\Im[\omega]/\Omega_c=0.0109$, $t_0 = 160 P_c$ for $m=1$
and $\Im[\omega]/\Omega_c=0.0168$, $t_0 = 50 P_c$  for $m=2$; model II
are $\Im[\omega]/\Omega_c=0.00557$, $t_0 = 170 P_c$ for $m=1$ and
$\Im[\omega]/\Omega_c=0.0179$, $t_0 = 50 P_c$ for $m=2$; model III are
$\Im[\omega]/\Omega_c=0.00366$, $t_0 = 120 P_c$ for $m=1$ and
$\Im[\omega]/\Omega_c=0.0175$, $t_0 = 40 P_c$ for $m=2$; and model IV
are $\Im[\omega]/\Omega_c=0.00909$, $t_0 = 40 P_c$ for $m=1$ and
$\Im[\omega]/\Omega_c=0.00459$, $t_0 = 150 P_c$ for $m=2$.  Therefore,
a clear relation between the dominancy of the saturation amplitude of
$M_m$ diagnostics and the strength of the characteristic complex
frequency can be seen in Fig. \ref{fig:dig}.  There seems to be a
threshold between $\Gamma = 1.33$ and $1.5$ to change the dominancy of
the $m$ mode.  This feature can roughly be understood from the
strength of the instabilities by investigating the imaginary part of
the eigenfrequency.  Evaluating the amplification timescale derived in
Eq.~(\ref{eqn:timescale_ref}) for each eigenfrequency computed in
Tables~\ref{tab:qnm_cyp150-1}, \ref{tab:qnm_cyp150-2},
\ref{tab:qnm_cyp150-3}, and \ref{tab:qnm_cyp150-4}, the imaginary part
of the reflection eigenfrequency has a clear dependence on the
stiffness of the equation of state.  We summarize our finding in
Fig. \ref{fig:dominancy}.

\section{Conclusions
\label{sec:Conclusions}}
We have investigated the unstable features of low $T/W$ dynamical
instabilities in differentially rotating stars in terms of a wide
range of the stiffness of the equation of state.  We have adopted a
normal mode analysis and a scattering rising from the corotation
barrier in the equatorial plane, and compare the results with those of
three-dimensional hydrodynamic simulations.

Unstable normal modes for low $T/W$ dynamically unstable stars are
found in the linear analysis, and they are qualitatively confirmed by
an amplified oscillation of the scattering sound waves between
corotation and the surface.  Although the growth timescale is in
agreement on a qualitative level, the criterion has clear agreement
with the results of both numerical simulations and normal mode
analyses.  We do not find any additional modes to the well-known $f$
and $p$ modes in the linear analysis for both stable and unstable
stars, but the stability of the system may change when the corotation
barrier appears in the effective potential.  The resonant frequency in
both cylindrical and spheroidal models in the linear analyses agrees
with that of hydrodynamic simulations when the deformation rate of the
rotating configuration approaches zero (non-rotating configuration).
The above fact confirms our models to be efficient for finding low
$T/W$ dynamically unstable stars.

\begin{figure*}
\centering
\includegraphics[keepaspectratio=true,width=16cm]{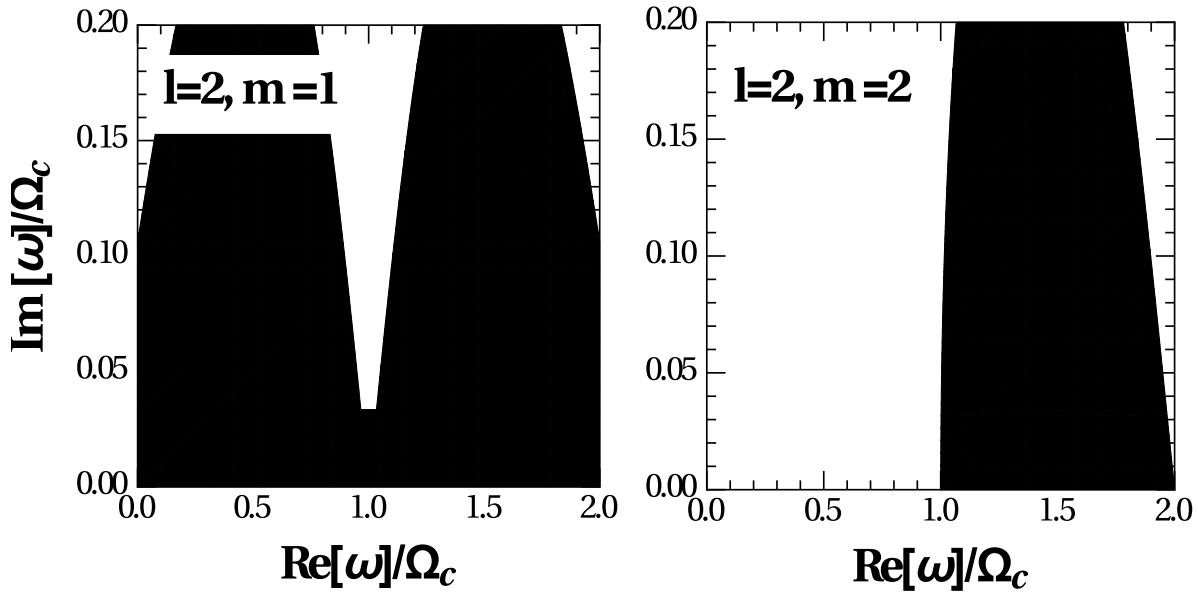}
\caption{
Allowed frequency region (white region) to treat the pulsation system
for spheroidal models as an eigenvalue problem.  The left panel
represents the case of $l=2$, $m=1$, and the right one represents that
of $l=2$, $m=2$.
\label{fig:FreqRegion}
}
\end{figure*}

The eigenfunction of the modes is also found to display a similar
behavior to the well known $f$ and $p$ modes.  Once corotation exists
inside the star, the perturbed enthalpy oscillates between corotation
and the surface.  This may indicate that the perturbed enthalpy is
affected by the corotation barrier, and therefore cannot cross
corotation.  This feature requires reinterpretation of the pulsation
modes in rotating stars when a corotation singularity exists inside
the stars.

Finally we are able to constrain the stiffness of the equation of
state by the direct observation of mode decomposed gravitational waves
from low $T/W$ dynamically unstable stars.  Investigating the
dominancy of the azimuthal mode in the normal mode analysis in a
cylindrical model, the threshold of the $m=2$ bar mode is around
$\Gamma \approx 1.50$.  Using the above fact, we are able to constrain
the stiffness of the equation of state by focusing the ratio between
$m=1$ and $m=2$ of the gravitational waveform.

We have computed the linear analysis in the equatorial plane to reduce
the basic pulsation equations to the ordinary differential ones.  Our
results clearly show that a rotational configuration of the star
should be fully taken into account.  In order to achieve complete
agreement between the linear analysis and hydrodynamic simulation, a
two-dimensional eigenmode analysis with corotation duly considered is
required, which is a challenging task in this field.

\begin{acknowledgments}
This work was supported in part by JSPS Grant-in-Aid for Young
Scientists B (No.~23740201), Grant-in-Aid for Scientific Research B
(No.~16H03986), and by the Waseda University Grant for Special
Research Projects (2014K-6100).  Numerical computations were performed
on the Cray XC40 cluster in the Yukawa Institute for Theoretical
Physics, Kyoto University, on the Cray XC30 cluster in the Center for
Computational Astrophysics, National Astronomical Observatory of
Japan, and on the cluster at Relativistic Astrophysics Group at the
Research Institute for Science and Engineering, Waseda University, and
at High Energy Astrophysics Group at Department of Physics, Waseda
University.
\end{acknowledgments}

\appendix*
\section{Boundary condition at center in spheroidal models
\label{chap:append_a}}
We adopt the technique of \citet{Unno89} for imposing a regularity
condition at the center.  The basic pulsation equations at the center
can be written as
\begin{equation}
r \frac{d}{dr}
\left[
\begin{array}{c}
\delta U_{lm} \\
\chi_{lm} \\
\delta \Phi_{lm} \\
\psi_{lm}
\end{array}
\right]
=
\left[
\begin{array}{cccc}
0 & 1 & 0 & 0\\
\alpha_{lm} & \beta_{lm} & 0 & 0\\
0 & 0 & 0 & 1\\
0 & 0 & l(l+1) & -1
\end{array}
\right]
\left[
\begin{array}{c}
\delta U_{lm} \\
\chi_{lm} \\
\delta \Phi_{lm} \\
\psi_{lm}
\end{array}
\right]
,
\label{eqn:Matrix_c}
\end{equation}
where $\chi_{lm} = r (d \delta U_{lm} / dr)$, $\psi_{lm} = r (d \delta
\Phi_{lm} / dr)$,
\begin{eqnarray}
\alpha_{lm} &=& l (l+1) - q_{lm} [ l(l+1) - m^2], \\
\beta_{lm} &=& -1 + q_{lm},\\
q_{lm} &=& \frac{4 \Omega_c^2}{\tilde{\omega}^2}.
\end{eqnarray}
The four eigenvalues of the matrix in Eq.~(\ref{eqn:Matrix_c}) are
\begin{equation*}
\lambda_{lm} = \lambda_{lm}^{(1)}, \quad \lambda_{lm}^{(2)}, \quad l,
\quad -(l+1),
\end{equation*}
where $\lambda_{lm}^{(1)}$ and $\lambda_{lm}^{(2)}$ satisfy
\begin{equation}
\lambda_{lm}^2 - \beta_{lm} \lambda_{lm} - \alpha_{lm} = 0.
\label{eqn:lambda}
\end{equation}
It is clear from the matrix in Eq.~(\ref{eqn:Matrix_c}) that the real
parts of the eigenvalues $\lambda_{lm}^{(1)}$ and $\lambda_{lm}^{(2)}$
correspond to the powers of $\delta U_{lm}$ and $\chi_{lm}$ at the
center, while $l$ and $-(l+1)$ correspond to those of $\delta
\Phi_{lm}$ and $\psi_{lm}$.

In order to close the system as an eigenvalue problem, only 2 out of 4
degrees of freedom at the center are needed since 1 freedom represents
scaling for the whole system, and the other freedom represents the one
of a surface boundary condition.  Since the power $-(l+1)$ for $\delta
\Phi_{lm}$ is already discarded because of the regularity condition at
the center ($l \agt 0$), only one of the powers $\lambda_{lm}^{(1)}$
or $\lambda_{lm}^{(2)}$ for $\delta U_{lm}$ should be discarded.  The
above condition can be written in general as 
\begin{equation}
\max (\Re[\lambda^{(1)}_{lm}], \Re[\lambda^{(2)}_{lm}]) \geq 1, \quad
\min (\Re[\lambda^{(1)}_{lm}], \Re[\lambda^{(2)}_{lm}]) < 1.
\end{equation}
Therefore, we restrict the frequency regime with the above condition.
We show the allowed frequency region for the case of $l=2$, $m=1$ and
$2$ in Fig.~\ref{fig:FreqRegion}.

Finally, the regularity conditions at the center are written as
\begin{equation*}
\delta U_{lm} = C_{lm}^1 r^{\max(\Re[\lambda^{(1)}_{lm}], \Re[\lambda^{(2)}_{lm}])},
\quad
\delta \Phi_{lm} = C_{lm}^2 r^l,
\end{equation*}
where $C_{lm}^1$ and $C_{lm}^2$ are constants.

\bibliography{dominancy}

\end{document}